\documentclass[preprint2]{pp7}

\newcommand{\rsun}{$R_\odot$}
\newcommand{\lsun}{$L_\odot$}
\newcommand{\msun}{$M_\odot$}
\newcommand{\msunyr}{$M_\odot$\,yr$^{-1}$}
\usepackage{xcolor}
\usepackage{tabularx}
\usepackage{hyperref}
\input{pp7.h}

\begin{document}

\title{\textbf{\LARGE Accretion Variability as a Guide to Stellar Mass Assembly}}

\author {\textbf{\large William J. Fischer}}
\affil{\small\it Space Telescope Science Institute, 3700 San Martin Dr, Baltimore, MD 21218, USA }
\author {\textbf{\large Lynne A. Hillenbrand}}
\affil{\small\it Department of Astronomy, MC 249-17, California Institute of Technology, Pasadena, CA 91125, USA}
\author {\textbf{\large Gregory J. Herczeg}}
\affil{\small\it Kavli Institute for Astronomy and Astrophysics, Peking University, Yiheyuan Lu 5, Haidian Qu, 100871 Beijing, China}
\affil{\small\it Department of Astronomy, Peking University, Yiheyuan Lu 5, Haidian Qu, 100871 Beijing, China}
\author {\textbf{\large Doug Johnstone}}
\affil{\small\it NRC Herzberg Astronomy and Astrophysics, 5071 West Saanich Rd, Victoria, BC V9E 2E7, Canada\\ Department of Physics and Astronomy, University of Victoria, 3800 Finnerty Rd, Victoria, BC V8P 5C2, Canada}
\author {\textbf{\large \'Agnes K\'osp\'al}}
\affil{\small\it Konkoly Observatory, Research Centre for Astronomy and Earth Sciences, E\"otv\"os Lor\'and Research Network (ELKH),\\ Konkoly-Thege Mikl\'os \'ut 15-17, 1121 Budapest, Hungary\\CSFK, MTA Centre of Excellence, Konkoly-Thege Mikl\'os \'ut 15-17, 1121 Budapest, Hungary\\Max Planck Institute for Astronomy, K\"onigstuhl 17, 69117 Heidelberg, Germany\\ELTE E\"otv\"os Lor\'and University, Institute of Physics, P\'azm\'any P\'eter s\'et\'any 1/A, 1117 Budapest, Hungary}
\author {\textbf{\large Michael M. Dunham}}
\affil{\small\it Department of Physics, State University of New York at Fredonia, 280 Central Ave, Fredonia, NY 14063, USA}

\begin{abstract}
\baselineskip = 11pt
\leftskip = 1.5cm
\rightskip = 1.5cm
\parindent=1pc
{\small
Variable accretion in young stellar objects reveals itself photometrically and spectroscopically over a continuum of timescales and amplitudes.  Most dramatic are the large outbursts (e.g., FU~Ori, V1647~Ori, and EX~Lup type events), but more frequent are the less coherent, smaller burst-like variations in accretion rate. Improving our understanding of time-variable accretion directly addresses the fundamental question of how stars gain their masses. We review variability phenomena, as characterized from observations across the wavelength spectrum, and how those observations probe underlying physical conditions. The diversity of observed lightcurves and spectra at optical and infrared wavelengths defies a simple classification of outbursts and bursts into well-defined categories. Mid-infrared and submillimeter wavelengths are sensitive to lower-temperature phenomena and more embedded, younger sources, and it is currently unclear if observed flux variations probe similar or distinct physics relative to the shorter wavelengths. We highlight unresolved issues and emphasize the value of spectroscopy, multiwavelength studies, and ultimately patience in using variable accretion to understand stellar mass assembly.
 \\~\\~\\~}
\end{abstract}

\section{\textbf{INTRODUCTION}}

Young stellar objects (YSOs) vary both photometrically and spectroscopically on a range of timescales and amplitudes. While some variability\index{variability} patterns trace the rotation of the young active star or the effects of intervening dust, changes in overall luminosity are mainly due to changes in the rate of accretion\index{accretion} from the disk to the star that are caused by temporal changes in the star-disk connection and in the flow of gas through the disk. It is not always easy to disentangle the signatures of variable accretion from those of other effects such as changing extinction, even with multiwavelength observations. Still, by attempting to probe variable accretion rates, observers are beginning to answer the fundamental question of how stars gain their masses.

\begin{figure*}[!ht]
\begin{center}
 \includegraphics[width=\textwidth]{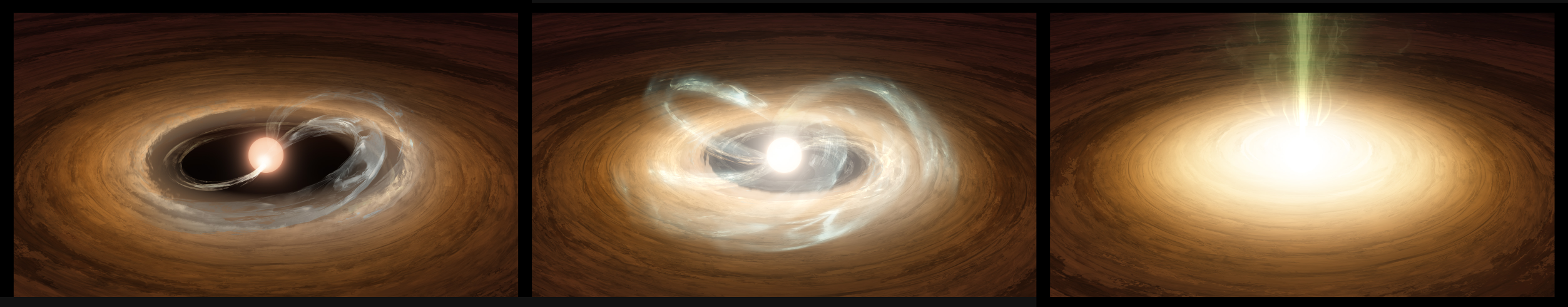}
 \caption{\small Artistic demonstration of how an accreting disk transitions from routine magnetospheric accretion to outburst accretion, accompanied by a rise in luminosity. Credit: T. Pyle (Caltech/IPAC). 
 }
 \label{fig:cartoon}
\end{center}
\end{figure*}

Understanding how accretion rates vary is important in the protostellar phase, when the stellar mass is mostly built up, as well as in the pre-main sequence (T Tauri) phase, when accretion rates from the disk are relatively small but important to disk evolution and planet formation. Although accretion bursts occur with durations as short as hours, large but rare accretion outbursts\index{outbursts} that last for months to decades play an outsized role in stellar growth and disk evolution. There are also potentially long-lasting consequences for the disk chemistry and for the evolution of the stellar radius. While the more common accretion variability with moderate duration and amplitude may not significantly influence stellar mass assembly, its observability still provides a quantifiable window to the underlying physics of disk accretion.  In this chapter, we review variable accretion, with an emphasis on outburst behavior but also including lower-amplitude phenomena that can be characterized as routine accretion variability. We focus on observational diagnostics that constrain disk accretion physics and the implications of these phenomena for stellar mass assembly over the first few Myr of stellar evolution. 

Some of the questions currently at the forefront of research on stellar mass assembly are: What is the role of variable accretion in building stellar mass and determining the observed properties of protostars? Do all stars experience large-amplitude accretion outbursts as they form, or do these require special circumstances? What triggers large accretion events? What switches them off or temporarily stops them? How can variability be used to evaluate the instabilities that lead to bursts? How well are we measuring ``accretion luminosity'' from monochromatic observations? How does variability influence chemical or mineralogical processes that affect the formation of planets?

A wide range of variably accreting sources with diverse and sometimes unique properties has been discovered, particularly in the years since Protostars \& Planets VI. At that time \citep{audard14}, there was emphasis on two major classes: The FU~Ori stars as major outbursts\index{outbursts} that last for decades to centuries and may or may not recur, and the EX Lup stars as less energetic outbursts that last for months to about a year and are observed to recur. Ongoing research shows that the peak luminosity and subsequent evolution of the lightcurve vary significantly among members of each class. Additionally, many outbursting YSOs have peculiar properties and do not fit into these two classical categories, in terms of their lightcurves or spectroscopic properties.

Outburst types are classified from lightcurves and/or spectra, but not always consistently. Even stars that are broadly accepted to be FU~Ori outbursts cover a wide range of luminosities, spanning two or more orders of magnitude.  We discuss and evaluate the (f)utility of the widespread practice of assigning accretion bursts into a small number of seemingly non-distinct categories. This is a particularly important and timely exercise as the parameter space over which outbursts are detected is expanding, e.g., in time, in survey volume, and in wavelength. 

We begin in \S~\ref{sec:basics} by introducing the parameters we wish to constrain and what is learned from various diagnostics. In \S~\ref{sec:big}, we discuss the ability to understand stellar mass assembly from analysis of protostar luminosities, concluding that we must look beyond them to distinguish among accretion models.

In \S~\ref{sec:acc}, we review what is known about photometric and spectroscopic variability and its impact on mass accretion. We discuss typical low-state accretion and stochastic events in pre-main sequence stars and protostars\index{protostars} and review the literature for large outbursts. We close the section with a discussion of indirect signatures of outbursts.

In \S~\ref{sec:unresolved}, we focus on unresolved issues. We discuss theoretical notions as to what causes accretion variability and what ends outbursts, consider evidence for distinct characteristics of FU Ori type disks, and note that accurate outburst rate determination is challenging. We close the section with a discussion of best practices for follow-up classification. We summarize the review in \S~\ref{sec:summary}.

\section{\textbf{SETTING THE SCENE FOR STUDIES OF MASS ASSEMBLY}}\label{sec:basics}

Stars acquire mass initially via the roughly spherical collapse of a fragmented molecular cloud core to form a protostar, and then through accretion via a disk that persists well into the optically visible pre-main sequence evolutionary phases of the central source. The accretion process at all stages is observed to be variable in time, as traced by time-series observations, on a wide range of timescales. The relative importance of these different accretion states in building up the central stellar mass, however, is currently poorly understood. To set the scene for this review, Figure~\ref{fig:cartoon} illustrates how an accreting disk transitions from routine magnetospheric accretion\index{magnetospheric accretion} to outburst accretion and is accompanied by a rise in luminosity.

We begin with an overview of what we aim to constrain when observing accreting YSOs. Observational studies use complex techniques to extract as much information as possible from limited data sets, so we first establish what parameters are most important to measure. We then take a look at what is learned from various diagnostics, beginning with the stellar magnetosphere and moving outward to the circumstellar envelope.

\subsection{\textbf{Physical Parameters to Be Constrained}}\label{sec:relevant}

Given the fundamental stellar parameters of mass $M_*$\index{stellar mass} and radius $R_*$\index{stellar radius} and the evolution of the mass in time under the influence of disk accretion, $\dot M_{\rm acc}$\index{accretion rate}, the resulting system luminosity $L_{\rm proto}$, the sum of the photospheric and accretion luminosities, can be expressed as
\begin{equation}
L_{\rm proto} = L_{\rm phot} + L_{\rm acc} = L_{\rm phot} + G M_* \dot M_{\rm acc}/R_*\label{e.lum},
\end{equation}
assuming that all accretion energy is radiated away. Clearly, this quantity evolves in time. A central theme of our discussion is thus: ``How do stars acquire the masses we believe they have, on timescales corresponding to the ages we believe they are, with the luminosities at which they are observed?'' In order to address this question, we need to consider what variables actually matter, and then consider how we can constrain them and their evolution through observables.

From the equation above we have radiative parameters such as $L_{\rm phot}$ and $L_{\rm acc}$, which relate to the observables. Although important in determining $L_{\rm acc}$, the factor $M_*/R_*$ varies by only factors of several as a function of $M_*$, as well as over time during the late protostellar and early pre-main sequence evolutionary phases. Furthermore, for protostars\index{protostars} we are generally in a position of not having good estimates of $M_*$, $R_*$, or $M_*/R_*$ beyond reasonable assumptions. So although these variables do change on long timescales as a result of accretion, and in the case of $R_*$ also normal pre-main sequence contraction, they are not our focus here.

The disk property $\dot M_{\rm acc}$ is key, with other disk parameters such as its mass $M_{\rm disk}$\index{disk mass} and radius $R_{\rm disk}$\index{disk radius} also of clear importance for understanding the mass reservoir that is available for accretion. There is no a priori reason to expect that the disk efficiently processes accreting material, either that falling in from an envelope \citep{kuffmeier18}, or that moving radially inward due to viscous\index{viscous evolution} accretion {and other disk instabilities} \citep{bai11}.  The potential mismatch \citep[as observed, for example, between envelope infall and disk accretion rates by][]{kospal17b} suggests that instabilities are to be expected. They could be regularly occurring and possibly periodic, or irregularly occurring as discrete episodic events \citep{armitage15}.

Finally, as we are interested in understanding the accumulation of stellar mass\index{mass assembly} and the role of variable mass accretion history in that process, there is also of course time $t$. As in many areas of astronomy, we can consider populations of young stellar objects to follow the distribution of accretion properties as a function of estimated population age. However, we also have the fortune of being able to observe variable accretion in action, with an increasing number of systems observed to undergo accretion rate changes. These span the range from detailed monitoring of the hour-to-day-to-week standard accretion processes \cite[e.g.,][]{sousa16,manara21,robinson19}, to large-scale time-domain surveys (Gaia; ASAS-SN; ZTF; ATLAS) that are capable of detecting outbursts in the act, and following their evolution from quiescence to outburst and back to low-state accretion again.

In order to understand the relevance to stellar mass assembly of the different accretion modes, we need to constrain parameters beyond instantaneous measurements of presumed steady-state or time-averaged accretion rates.  Specifically, we also need to measure burst and outburst amplitudes, peak accretion rates, durations, and repeat cadences, but most importantly $\int \dot M_{\rm acc} ~dt$ per event, for all events. To probe the physics operating within accretion disks we need detailed lightcurves that include rise shapes and decay shapes, as well as spectroscopic sampling at key phases. Spectra add further precision to our understanding of the outburst physics by quantifying disk temperature, which can be used to quasi-independently estimate $\dot M_{\rm acc}$. Long-wavelength and spatially resolved observations can also constrain the reservoir disk mass, radial size, and vertical thickness, as well as detailed spatial structure.

The recent observational literature on variable accretion can be described along four main avenues: characterization of typical variability types, timescales, and amplitudes as a function of observed wavelength;  monitoring over years to decades of large samples of YSOs for statistical studies of outburst frequency; dedicated searches for recent burst and outburst candidates among the sea of variable YSOs (and other astronomical objects); and careful quantitative measurements of the properties of individual sources known to be in an outburst state. Such a mix of studies is necessary to further our understanding of accretion outburst physics.

\subsection{\textbf{Diagnostics of Accretion Heating and Cooling}}

Changes in the accretion rate may be measured through a variety of diagnostics.  Direct diagnostics of accretion (see Chapter by Manara et al.\ in this volume) involve measurements of heated gas that is shocked as it decelerates and merges with the stellar photosphere. The primary diagnostics are the ultraviolet continuum and various emission lines throughout the ultraviolet and optical, which are measurable in low-extinction cases with favorable system geometry; proxy lines in the red optical and infrared are used for more embedded sources. In other cases, we are prevented from detecting the star directly and must rely on reprocessed emission that emerges at infrared or millimeter wavelengths. This section steps through the different physical components of an accreting protostar and how they are detected, to set the stage for how accretion variability\index{variability} is measured.

\textbf{Magnetospheric accretion:}\index{magnetospheric accretion} For classical T Tauri stars, accreting gas is channeled by stellar magnetic fields from the disk onto the star. The accretion shock itself is mostly or entirely buried beneath the photosphere, with X-rays heating the nearby photosphere. The energy from the heated photosphere and the accretion funnel flow produces hydrogen recombination and $H^-$ continuum, along with line emission, both best measured at optical and ultraviolet wavelengths. Inner disk warps, often associated with the magnetospheric flow, may block the stellar light and/or radiation from the accretion flow. Accretion rates are measured through either modeling of continuum excess emission, requiring detailed knowledge of stellar parameters,  accretion geometry, and temperature structure, or via proxy emission line diagnostics where extinction-corrected line fluxes are calibrated to the accretion shock models  (see reviews by \citealt{hartmann16} and Manara et al., in this volume). Changes in accretion can be tracked through changes in continuum brightness or in line fluxes.

\textbf{The passively heated dust disk:}\index{passive heating} The dust temperature in the disk declines from the dust sublimation temperature of $\sim 1400$ K at small radii to $\sim 10-20$ K at large radii. For low-accretion disks, the heating is passive, with the reprocessing of starlight at the disk surface dominating over the energy liberated locally by the diffusion of material through the disk. The disk temperature is thus highest at the surface.  The innermost dust is warmest and emits in the near and mid infrared. The outer disk is cold and dominates the long-wavelength emission. 

\textbf{The viscously heated gas disk:}\index{viscous heating} For rapidly accreting disks, viscous processes heat the disk to temperatures that can be higher than the star's, over areas much larger than the star. The system luminosity thus becomes dominated by the accretion. If the accretion rate is high enough, the accretion flow in the innermost disk region may overwhelm the magnetic pressure and crush the magnetosphere, disrupting the magnetospheric flow. Viscously heated disks are most commonly measured at optical and infrared wavelengths, with shorter wavelengths tracing hotter disk material close to the star. Accretion rates are measured through models that make the simple assumption that luminosity correlates with accretion rate. Transitions from a typical low-state accretion disk to a high-state viscously heated disk are seen as large increases in brightness, typically detected at optical and infrared wavelengths. Changes in brightness can be directly converted to changes in accretion rate, once possible accompanying changes in extinction are taken into account.

\textbf{Envelope emission:}\index{envelope emission} In the case of protostars\index{protostars}, the bulk of the inner envelope mass is heated by the star, while the exterior is heated by the interstellar radiation field. Since the envelope does not extend as close to the star as the disk does, the envelope is cooler than the disk and emits at longer wavelengths, with a peak in the far IR. The large dust reservoir acts as a bolometer, and thus changes in the protostellar luminosity, e.g., due to accretion from the disk onto the star, vary the degree of dust heating and hence the wavelength response of the re-emission. Accretion rates are measured through radiative transfer modeling of spectral energy distributions.  Changes in protostellar accretion rates may thus be detected via changes in long-wavelength emission from the envelope.  

\section{\textbf{STELLAR MASS ASSEMBLY AS PROBED BY LUMINOSITIES}}\label{sec:big}

In this section, we review what is known about the luminosities of protostars and pre-main sequence stars and how to interpret them. We use the term ``protostar'' to refer to YSOs (including the central protostellar object and a disk, if already formed) that are still embedded in envelopes with significant envelope infall and accretion rates (the so-called Class 0 and Class I objects), whereas we use the term ``pre-main sequence star'' to refer to YSOs (and associated disks) that are beyond the embedded phase of evolution, with at most only residual envelopes remaining (the so-called Class II and Class III objects).

\begin{figure}[!t]
\includegraphics[width=\columnwidth]{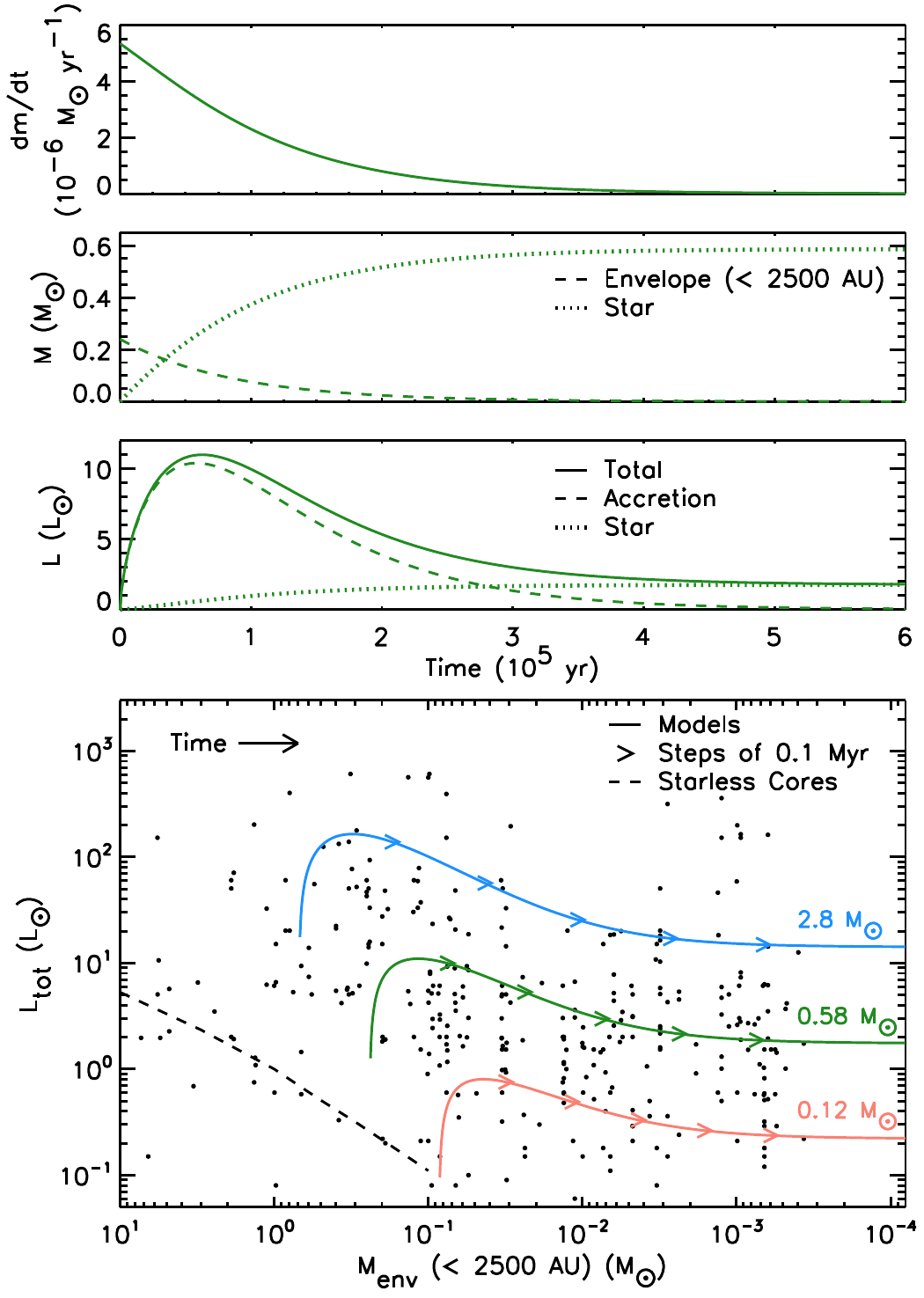}
  \caption{
  Representative example of the connection between accretion and luminosity during the protostellar stage of evolution, adapted from \citet{fischer17}. The top three panels show how the infall rate; envelope mass inside 2500 au and stellar mass; and total, accretion, and stellar luminosities change over time for a model featuring exponentially declining infall from a protostellar envelope that forms a star of mass 0.58 $M_\sun$. The bottom panel plots luminosity against envelope mass inside 2500 au derived from fitting the 1--870 \micron\ SEDs of 324 Orion protostars. The dashed curve shows luminosities for externally heated spherical starless cores ranging in mass from 0.1 to 10 $M_\sun$. Solid curves show the luminosity evolution of envelopes with exponentially declining infall rates from a selection of initial masses. From bottom to top, the final stellar masses are 0.12, 0.58, and 2.8 $M_\sun$. Time increases from left to right, with arrowheads marking times from 0.1 to 0.5 Myr in steps of 0.1 Myr. Stellar properties are based on fits to the model tracks of \citet{siess00}. }\label{fig:lum}
\end{figure}

Figure~\ref{fig:lum} shows the connection between accretion and luminosity during the protostellar stage of evolution, using observations and analytic models from \citet{fischer17}.  For the envelope infall rate illustrated in the first panel, the second panel shows how the mass of the envelope within 2500 au of the star decreases and the mass of the star increases over time. The third panel shows how, for this model, the accretion luminosity peaks early and falls off with time (Equation~\ref{e.lum} above), while the stellar luminosity slowly increases as the star gains mass. The figure is meant only as illustrative of the connections between masses, mass accretion rates, and luminosities.

In an analysis of protostars\index{protostars} identified in the Taurus-Auriga region, \citet{kenyon90} found that protostars are underluminous, in a statistical sense, compared to expectations based on simple accretion arguments \citep[see also][]{kenyon94,kenyon95}. The popular phrase ``luminosity problem''\index{luminosity problem} originally referred to this discrepancy. The articulation of the issue was that for an assumed star formation time of 10$^5$\,yr, an average accretion rate of $\dot M_{\rm acc} = 10^{-5}$\,\msunyr\ is required to form a 1\,\msun\,star. Adopting $M_*=0.5$\,\msun\ as the mean mass of a 1\,\msun\ star over its formation, assuming a constant accretion rate, and further adopting $R_*=3$\,\rsun\ and $\dot M_{\rm acc}=10^{-5}$\,\msunyr, Equation~\ref{e.lum} gives $L_{\rm proto} \gtrsim 50$\,\lsun, where the inequality acknowledges that including photospheric luminosity would only increase this value. However, \citet{kenyon90} found the observed luminosities of protostars in Taurus-Auriga to be typically around 1\,\lsun\ and thus underluminous compared to expectations. Two of the possible resolutions proposed by \citeauthor{kenyon90} include longer star formation times and episodic accretion, where long periods of slow accretion are punctuated by short bursts of rapid accretion.  For the latter solution, averaged over their lifetimes protostars would still feature the requisite high mean accretion rates but would typically be observed during their low-accretion states.

Revised measurements of protostellar lifetimes are indeed longer than the 10$^5$\,yr assumed by \citeauthor{kenyon90}  Results from {Spitzer Space Telescope}\index{Spitzer} mid-infrared\index{mid-infrared} surveys of nearby star-forming regions give protostellar lifetimes of $\sim$\,0.5\,Myr \citep{evans09,dunham14,dunham15}, $\sim$\,5$\times$ longer.  Although there remains some debate on the validity of these longer ages \citep[see, e.g.,][]{kristensen18}, if the 0.5 Myr lifteimes for envelopes are correct, they reduce the required accretion rates by the same factor, thus reducing the predicted protostellar luminosities to $L_{\rm proto} \gtrsim 10$~\lsun.

These surveys and others have consistently shown mean luminosities of samples of protostars to be a few \lsun\ (with medians of about 1 \lsun); extinction corrections tend to increase the luminosities. Thus the original discrepancy between predicted and observed luminosities of a factor of 50 or more is now reduced to a factor of 2 to 10, within the uncertainties from both observations and assumptions. Based on this reduction, \citet{offner11} claimed that the original luminosity problem has been resolved.

We now consider episodic accretion, the second resolution proposed by \citeauthor{kenyon90} The models with constant accretion rates that motivated the identification of the luminosity problem cannot reproduce the observed distribution of protostellar luminosities, but a variety of other accretion scenarios have been shown to be plausible. Reviewing the state of these attempts for Protostars \& Planets VI, \citet{dunham14} showed that accretion scenarios that spend much of their time at relatively low accretion rates, and less time at relatively high accretion rates, provide good matches to the observed luminosity distributions. These scenarios include models with accretion rates that generally decline with time, increase with instantaneous and/or final protostellar mass, and/or are dominated by large-amplitude stochastic fluctuations; i.e., episodic accretion \citep[e.g.,][]{myers09,myers10,offner11,dunham12}.

Since Protostars \& Planets VI, there has been relatively little work that explicitly addresses this topic. The finding that a broad range of time-dependent accretion rates, not necessarily episodic in nature, can match observed protostellar luminosities was further reinforced by \citet{fischer17}. They found that protostellar luminosities in Orion span a similarly large range of values at each evolutionary stage, but with median values that decline from 2.3\,\lsun\ for Class 0 protostars to 0.87\,\lsun\ for Class I protostars. They showed that this decline in luminosity coupled with a large spread can be explained by envelope masses and accretion rates that decline exponentially with time to form a range of stellar masses (Figure~\ref{fig:lum}). This result that episodic accretion is not necessarily required matches those from the theoretical studies cited in the previous paragraph, except now with a large sample of protostars at each evolutionary stage and a more complete comparison with model luminosities that removes some of the uncertainty over the validity of earlier results.

Protostellar luminosities and their dependence on evolutionary state are now well characterized but do not, by themselves, definitively distinguish among different accretion scenarios. Accretion scenarios with short-term, large-amplitude stochastic variability (possibly including episodic bursts), scenarios with long-term, small-amplitude secular variability, or scenarios with a combination of both have all been shown to provide viable matches to the observed luminosities of protostars.

The finding that alternative time-dependent accretion models match observed luminosity distributions should not be taken as evidence that episodic accretion does not happen at all, or even that stellar growth is dominated by smooth accretion rather than large outbursts. Indeed, there is substantial empirical evidence for accretion variability, as will be discussed in this chapter. Instead, they show that episodic accretion is not specifically required to explain the observed luminosities of protostars, as there are alternative accretion scenarios that are equally effective at matching the observed luminosities of protostars.

To match observations, these families of models typically require protostellar mass distributions that differ from standard initial mass functions \citep{mckee10}. For example, it can be inferred from Figure~\ref{fig:lum} that \citet{fischer17} form stars with masses skewed to larger values than a typical IMF predicts. They suggest that this may be due to one or more of episodic accretion, poorly known $M_*/R_*$ ratios, or incompleteness to low masses at the distance to Orion. To distinguish among models that give good matches to luminosity distributions, it will be necessary to measure protostellar masses by fitting molecular line data from protoplanetary disks \citep[e.g.,][]{okoda18,tobin20}.

Improvements in sensitivity and sample size alongside more detailed modeling over the last three decades have mostly resolved the initial concern that the mean predicted and observed protostellar luminosities differ by large factors. Despite this apparent resolution of the original luminosity problem, there is still discussion in the current literature of a luminosity problem, but with a notably different meaning. The phrase now generally refers to the fact that, instead of simply being too low on average, protostellar luminosities span an enormous range that encompasses 3--4 orders of magnitude at both young (Class 0) and older (Class I) evolutionary stages. Further, there is a large population of very low luminosity objects\index{very low luminosity objects} with $L_{\rm proto} \lesssim 0.1$~\lsun. These details are challenging to explain with simple accretion scenarios that feature constant accretion rates \citep[e.g.,][]{enoch09,kryukova12,fischer17}, and the viability of various accretion scenarios proposed as resolutions remains uncertain.

We suggest that the field distinguish between the historical ``protostellar luminosity problem'' of predicted luminosities being too high and the current ``protostellar luminosity spread.'' This suggestion is motivated by the following considerations:

\begin{enumerate}

    \item Using the same phrase for two distinct issues has the potential to cause confusion in the literature;

    \item Updated measurements of protostellar luminosities and lifetimes have reduced the magnitude of the originally noted conflict between observations and expectations;

    \item Several models have been developed that provide good matches to the observed luminosity spreads;

    \item The problem now is to distinguish among these models. Which are correct, if any?

\end{enumerate}

Luminosities alone do not seem to provide enough diagnostic power to distinguish among these models. Further progress depends on testing the existing theories with measurements of other system properties, including protostellar masses, protostellar radii, and the accretion rates of more deeply embedded sources, to evaluate mass transport versus time.

\section{\textbf{OBSERVATIONS OF DISK ACCRETION}}\label{sec:acc}

This section, the heart of the review, discusses what we know about variable accretion. We synthesize variability\index{variability} studies during typical accretion and during bursts in pre-main sequence stars and in protostars. We close by discussing indirect tracers of variable accretion.

Table~\ref{t.nomenclature} provides an overview of the nomenclature used in this review. Entries above the horizontal line refer to conceptual categories of variability. Entries below the line refer to observed phenomena that correlate to varying degrees with the conceptual categories. Figure~\ref{fig:amp_vs_time} plots amplitude versus timescale for various kinds of accretion-related variability as well as other types of variability for context.

\begin{table*}[!t]
     \centering
     \caption{Variability and Outburst Nomenclature}\label{t.nomenclature}
     \begin{tabularx}{\textwidth}{l|X}
Term & Explanation\\
\hline
Routine Variability\index{variability} & Changes typically $<$1--2 mag (factors of several) often accompanied by changes in emission line morphology and strength. In optical/near-IR, typical timescales are hours to days, similar to dynamical/rotational timescales of the star, magnetosphere, and very inner disk. Behavior may be quasi-periodic or stochastic and includes short-duration accretion flares and extinction dips. In mid-IR/sub-mm, variations over days to years are consistent with inner-disk viscous timescales\index{viscous evolution}. {\it Examples: most Class II classical T Tauri and Herbig Ae/Be stars, as well as Class I YSOs}.\\
\vspace{-0.2cm} & \\
Burst & Distinct, discrete brightness increase, typically $\sim$1--2.5 mag (factors of few to ten), with sustained but modest duration $\sim$1 week to typically $\lesssim$1 year. {\it Examples: DO Tau\index[obj]{DO Tau}, EX Lup\index[obj]{EX Lup} (2011).}\\
\vspace{-0.2cm} & \\
Outburst & Large brightness increase, typically $\sim$2.5--6 mag (factors of tens), taking months to years to develop, with a duration of many years to decades. {\it Examples: EX Lup\index[obj]{EX Lup} (2008), V1647~Ori\index[obj]{V1647 Ori}, FU~Ori\index[obj]{FU Ori}.}\\
\hline
EX~Lup event\index{EX Lup events} & Sometimes called EXors, these bursts have amplitudes of several mag, last for several months, and may recur after a few years. Spectra resemble T Tauri stars with atypically high accretion.\\
\vspace{-0.2cm} & \\
V1647 Ori event\index{V1647 Ori events} & These have amplitudes and durations as well as spectral characteristics intermediate between those of EX Lup bursts and FU~Ori outbursts\index{outbursts}.\\
\vspace{-0.2cm} & \\
FU~Ori event\index{FU Ori events} & Sometimes called FUors, these are outbursts with sustained increases in bolometric luminosity. Flux from the disk greatly exceeds flux from the star at optical wavelengths. They have long decay timescales similar to human lifetimes. A viscously heated disk dominates the spectrum, and the effective temperature associated with the spectrum decreases with increasing wavelength.\\
\vspace{-0.2cm} & \\
FUor candidate & Has an observed apparent outburst, but lacks spectroscopic confirmation of FU~Ori status. Some FUor candidates wind up being extinction clearing rather than accretion outbursts.\\
\vspace{-0.2cm} & \\
FUor-like & A source with a spectrum having the characteristic wavelength-dependent absorption features of FU~Ori stars, but no actual outburst was observed.\\
\hline
     \end{tabularx}
     \label{tab:nomenclature}
\end{table*}

\subsection{\textbf{Variable Accretion at Late and Early Times}}

Most stellar growth occurs when the protostar is still deeply embedded in its envelope. As gas in the disk flows to the star, the disk is quickly replenished by accretion from the envelope onto the disk.  However, the stages in the stellar growth that are most readily visible occur after the young star has shed its envelope and has reached nearly final mass. Measured accretion rates and disk survival timescales are consistent with 1--10\% of the final stellar mass accreted during this last stage of stellar assembly (see analyses by, e.g., \citealt{rosotti17}, \citealt{mulders17}, and \citealt{manara19}). Disk masses are typically less than 1\% of the stellar mass, though highly uncertain, and the disk can no longer access a larger reservoir of gas.

\begin{figure}[t]
\includegraphics[width=\columnwidth,trim={2cm 0.8cm 2cm 1.5cm},clip]{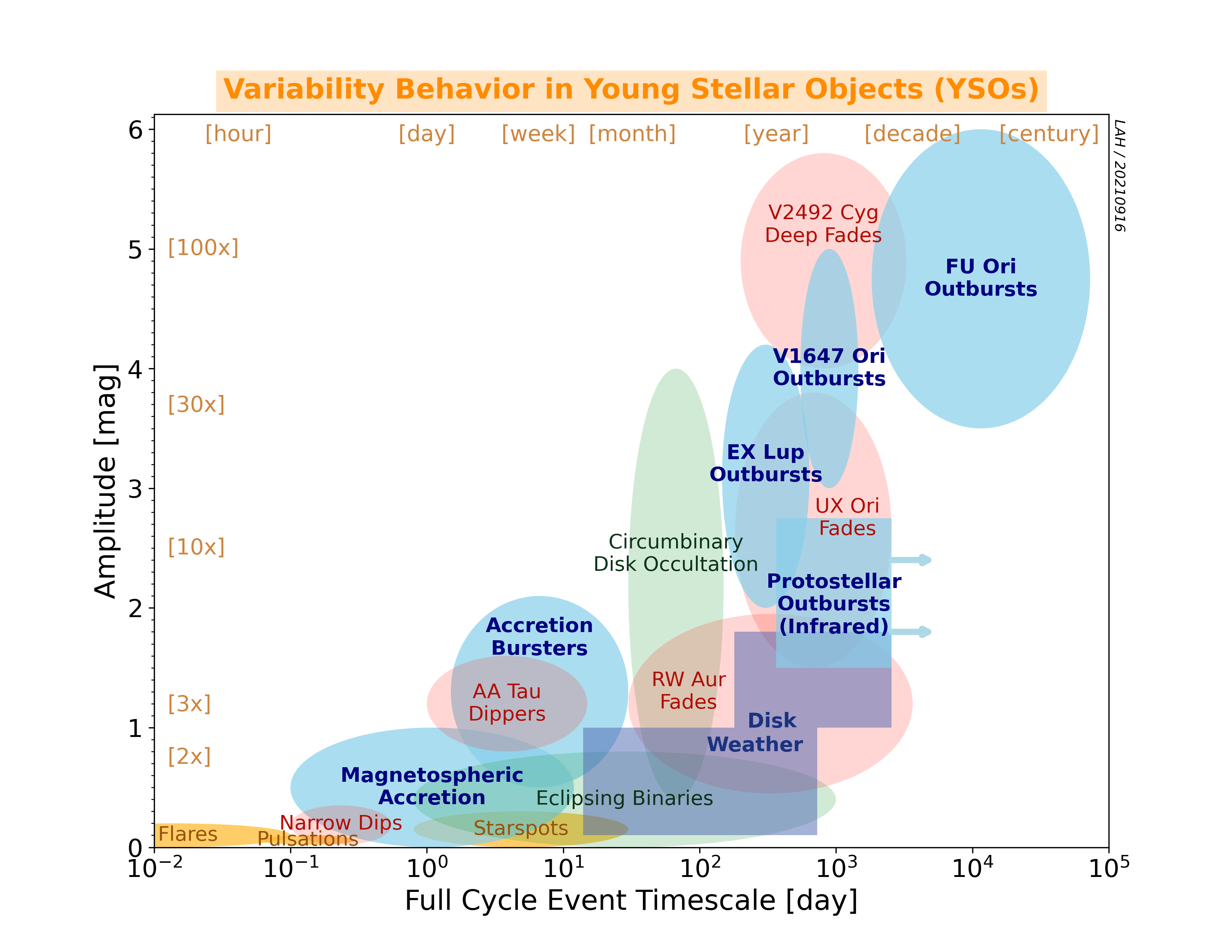}
\caption{\small 
Amplitude versus timescale for various flavors of YSO variability\index{variability}. Blue indicates the accretion-related events that are the focus of this review. Purple shows the routine variability, either brightening or fading, that is detected at longer wavelengths, which we refer to as ``disk weather". In addition, red indicates extinction-related behavior, yellow, stellar phenomena, and green, the variability expected from binary-related phenomena. We call attention to the overlap in this parameter space across different variability categories; boundaries are notional typical ranges.}\label{fig:amp_vs_time}
\end{figure}
 
In this section, we begin by describing  commonly observed changes in accretion at this very last stage of stellar growth. This variability\index{variability} is readily detected on timescales that are easy to probe and at visible wavelengths, where accretion is most directly measured. We then discuss the bursts and outbursts, events most commonly observed at or near the end stages of the main stellar accretion phase. The largest outbursts, {potentially for some YSOs} the last major event in stellar growth and disk draining, occur rarely and may persist for longer than a century. These descriptions establish the classes and physics of accretion variability (see also Table \ref{tab:nomenclature} and Figure \ref{fig:amp_vs_time}), which we extend to observations of younger sources where possible. We close with a consideration of the more challenging problem of measuring accretion variability during the main stage of stellar assembly, when dusty protostellar envelopes prevent the UV/optical studies that have enabled so much progress at later stages.
 
\subsubsection{Typical Low-State Routine Variability during Pre-Main Sequence Evolutionary Phases}\label{sec:obs}

\begin{figure*}[!t]
\begin{center}
\includegraphics[width=0.48\textwidth]{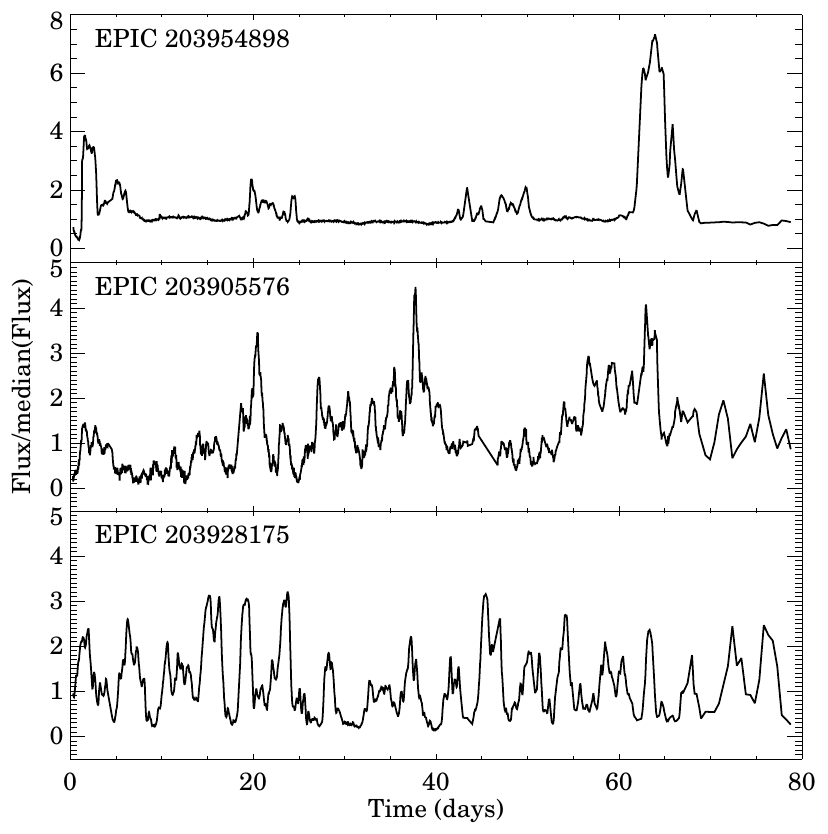}
\includegraphics[width=0.48\textwidth]{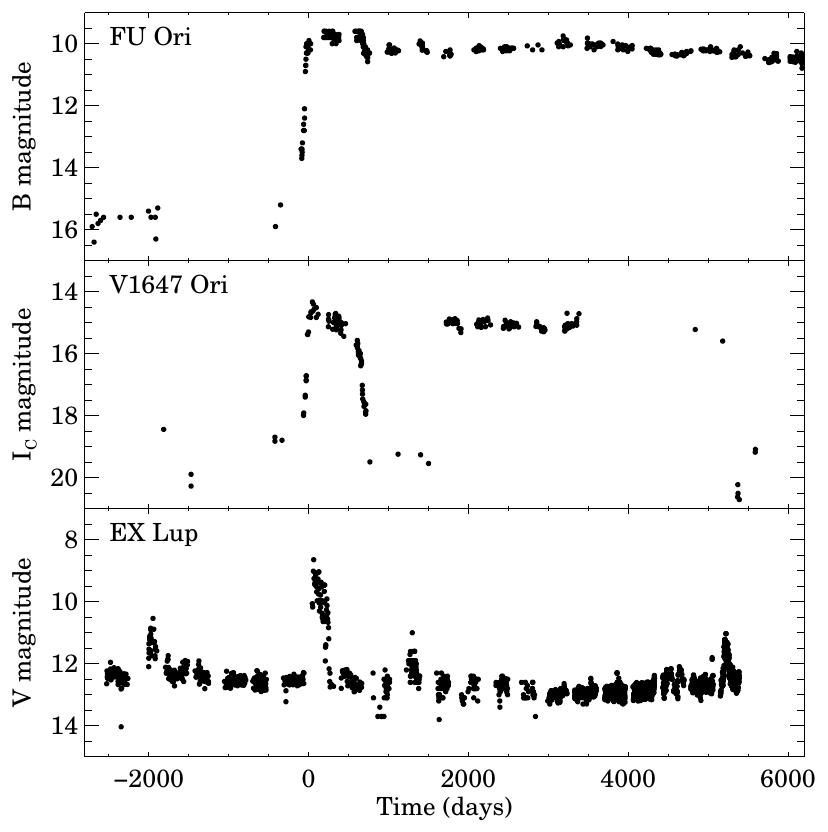}
\caption{\small Left panels: Typical low-state short-timescale accretion bursts in low-mass T Tauri stars \citep{cody17}. Right panel: Outburst lightcurves of prototypical objects (adapted from \citealt{kospal11}). See Table~\ref{t.nomenclature} for further context. $t=0$ corresponds to JD~$=$ 2,456,893 in the left panel and JD $=$ 2,428,600 for FU~Ori\index[obj]{FU Ori}, JD $=$ 2,453,000 for V1647~Ori\index[obj]{V1647 Ori}, and JD $=$ 2,454,440 for EX~Lup\index[obj]{EX Lup}.}\label{fig:lightcurves}
\end{center}
\end{figure*}

In the historical framework for interpreting typical variability behavior of accreting systems, established by, e.g., \citet{herbst94}, the extinction, accretion, and star spot coverage can all vary with time. Spots are periodically modulated by the stellar rotation, extinction often occurs in sharp drops that can be either short-lived or have a long duration, and accretion variability can be seen in sharp bursts that last hours to days, or as smooth changes on roughly weeklong to much longer timescales. Physical interpretations that attempt to characterize these phenomena from photometry alone are often degenerate.

In the past decade, space-based optical observatories (CoRoT\index{CoRoT}, MOST\index{MOST}, K2\index{K2}, and TESS\index{TESS}) have produced precise lightcurves from rapid (sub-hour) cadence monitoring\index{monitoring} extending over months, revealing new variability types and refining existing empirical variability classes. The highly structured lightcurves allow us to distinguish between accretion, spots, and extinction as the source of variability, and to infer accretion, star, and disk properties from the shape of those changes.

Variability in the magnetospheric accretion\index{magnetospheric accretion} process is mostly driven by how much mass is loaded from the disk onto the accretion flow, apparently a stochastic process. Discrete accretion bursts occur frequently, with amplitudes ranging from 10--700\% and durations of 1--10 days, repeated once every few to tens of days (\citealt{stauffer14,cody17}, see also Figure~\ref{fig:lightcurves}). In some cases, short bursts are seen on top of smooth changes over months. The brightness changes in these bursts are the consequence of stochastic instabilities\index{disk instabilities} in the star-disk connection \citep{stauffer16}. For close binaries, accretion pulses are triggered {by interactions between the stars and their disks} \citep{tofflemire17}, as expected from models \citep[e.g.,][]{munoz16}.

Changes in accretion rate tend to be dominated by short-timescale bursts coupled with rotational modulation. \citet{sergison20} found in $i$-band imaging that while most variability power is on the $\sim 1$ day timescale, consistent with the space-based lightcurves, some power is also found on $>10$ yr timescales. \citet{venuti15} found similar results with $u$- and $r$-band monitoring of NGC 2264\index[obj]{NGC 2264} over 7 years, with most power on short timescales. Likewise, \citet{rigon17} find that most accretors are dominated by low-level ($<0.3$ mag) variations on short timescales of days to weeks, consistent with rotational modulation; longer timescale excursions are typically larger and rare.

Spectroscopic\index{spectroscopy} monitoring\index{monitoring} campaigns typically show levels and timescales of variability similar to the photometry (e.g., \citealt{mendigutia13, costigan14,pouilly21}). Time-domain optical and ultraviolet spectroscopy, as well as X-ray lightcurves, are powerful for measuring the physical conditions and dynamics of the accretion flows \citep[e.g.,][]{alencar12,espaillat21,campbellwhite21,sousa21}. They thereby help us in interpreting accretion rate variations inferred from multi-band photometric variability
\citep[e.g.,][]{venuti21,zsidi22}. Evidence of correlation between optical photometric fluctuations and X-ray behavior related to accretion was provided by \cite{guarcello19}.

Spectroscopic monitoring with polarimetry\index{polarimetry} has revealed that magnetic accretion structures for fully convective low-mass stars are dipole-like \citep[e.g.,][]{alencar12}, though complicated by higher-order fields \citep{gregory11}. By combining K2 photometry with $ugri$ and H$\alpha$ monitoring of the Lagoon Nebula\index[obj]{Lagoon Nebula}, \citet{venuti21} found that the variability properties of low-mass stars are consistent with such dipole-like structures, while earlier-type stars have complex, higher-order fields that yield a smoother distribution of starspots. Models of lightcurves similarly indicate that dipole accretion structures tend to exhibit more flares or stochastic bursts that last for hours \citep{robinson21}. The interpretation of short-timescale variability signatures is generally consistent with the 3D MHD simulations \citep{romanova12,blinova16}, which predict stable and unstable regimes depending on the accretion rate and the strength and morphology of the magnetic field.  Simulations of accretion onto stars with weak magnetic fields may also produce variability \citep{takasao19}.

While the accretion measurements diagnose where the flow hits the star, when viewed at a nearly edge-on inclination ($i>70^\circ$) the connection between the funnel flow and the disk can be identified through brief dips in lightcurves. Stars that exhibit these are called dippers (or AA Tau-like stars). The funnel flow warps the base of the disk, and dust in this warp obscures the star as it rotates into our line of sight \citep{bouvier07}. These quasi-periodic dips are detected in $\sim 20-30$\% of lightcurves, last for hours to days, and are associated with changes in line-of-sight optical extinction of $A_V\sim 0.1-2$ mag \citep{cody10,stauffer15,roggero21}. Dippers are also identified in near-infrared \citep{rice15} and mid-infrared\index{mid-infrared} \citep{morales-calderon11} lightcurves. Simple models of the wavelength-dependent photometric effects include dusty dipolar accretion flows \citep{mcginnis15,kesseli16}.

For some stars, these extinction events are deep and do not repeat on a particular timescale, indicating the presence of large, unstable disk warps that quickly form and then disperse \citep[e.g.,][]{rodriguez17,guo18}. Some dippers have outer disks that are resolved with ALMA to have low inclinations, pointing to a misalignment between inner and outer disks \citep{ansdell16,loomis17,davies19}. Longer-term extinction events, or fades, are traditionally called RW Aur type systems, or UX Ori stars for objects that experience a blueing effect in their deep fades. These fades can persist for a decade or longer, with PV Cep\index[obj]{PV Cep} a well-known example, and AA Tau\index[obj]{AA Tau} a recent example. These extinction events may be caused by the rotational modulation of a warp at a large distance from the star \citep{bouvier13} or by an increase in the geometrical height of the inner disk (\citealt{covey21}; see also changes in mid-IR emission due to changing disk height in \citealt{espaillat11} and dynamics modeled by \citealt{flaherty16}). Dusty winds have also been suggested as a possible explanation for this behavior \citep{petrov15}.

The descriptions of these variable phenomena have been developed primarily from optical monitoring.  Recent infrared monitoring studies \citep[e.g.,][]{rice15, rebull15, wolk15, wolk18} confirmed that the short-timescale phenomena seen in optical lightcurves also manifest in the near- and mid-infrared\index{mid-infrared}, with clear bursters and dippers. Variability amplitudes were demonstrated to be larger for more embedded (protostellar) sources.  In addition, optical, near-infrared, and mid-infrared variability\index{variability} are not uniform.  Many objects show completely uncorrelated variability, including both high activity in the optical but low in the mid-infrared, or the inverse of high mid-infrared variability but only small changes in the optical \citep{morales-calderon11,cody14}. Dippers usually become redder in the dip,  but in some cases there is bluer-when-fainter behavior, perhaps a consequence of a change in inner disk structure and scattering \citep{mcginnis15}. The implications for the physics of variable accretion are unclear at present, but some optical variability appears to be exclusive to the hot accretion zone, while mid-infrared variability is exclusive to the disk and in many cases seems not to either influence or reflect the hot processes occurring closer to the star.

\subsubsection{Discrete Accretion Bursts and Outbursts}\label{s.gulps}

Beyond the standard routine accretion variability observed for most YSOs, described above, distinct brightening events in young stars come in a range of amplitudes, durations, and propensities to repeat. Figure~\ref{fig:amp_vs_time} plots amplitudes versus timescales, illustrating the significant increase in burst brightness (corresponding to increase in accretion rate) from EX~Lup to V1647~Ori to FU~Ori type outbursts\index{outbursts}; see also Table \ref{tab:nomenclature}. The observed diversity, however, defies classification into a small number of well-defined categories and may better reflect a continuum of empirical measurements. That said, there is a historical framework of classification that we use to frame our discussion. In what follows, we describe the lightcurves and spectra that typify each of these historical categories.

\textbf{EX~Lup bursts and outbursts:} Several decades ago, when the outburst phenomenon was first appreciated, objects with moderately large, 2.5--5\,mag optical outbursts that last for a few months to a few years, and repeat, were identified. After the prototype EX Lupi (and to rhyme with ``FUor''), these objects were named EXors \citep{herbig89}.

\looseness=1
The lightcurves show significant variety. EX~Lup\index[obj]{EX Lup}, for instance, while typically bursting at 1--3 mag, has displayed two more major $\sim$5\,mag outbursts, in 1955--56 and in 2008 \citep{abraham09,abraham19}. VY~Tau\index[obj]{VY Tau}, on the other hand, showed repeated 3--5\,mag outbursts for several decades until 1972, stayed in quiescence for 40 years, then resumed activity with $\sim$1.5\,mag bursts in 2012--2013 \citep{herbig90,dodin16}, and has been quiet since. Due to their extended durations and lengthy dormant periods, the duty cycle of EX Lup outbursts is not yet established.

\begin{figure*}[!t]
\begin{center}
\includegraphics[width=1\textwidth]{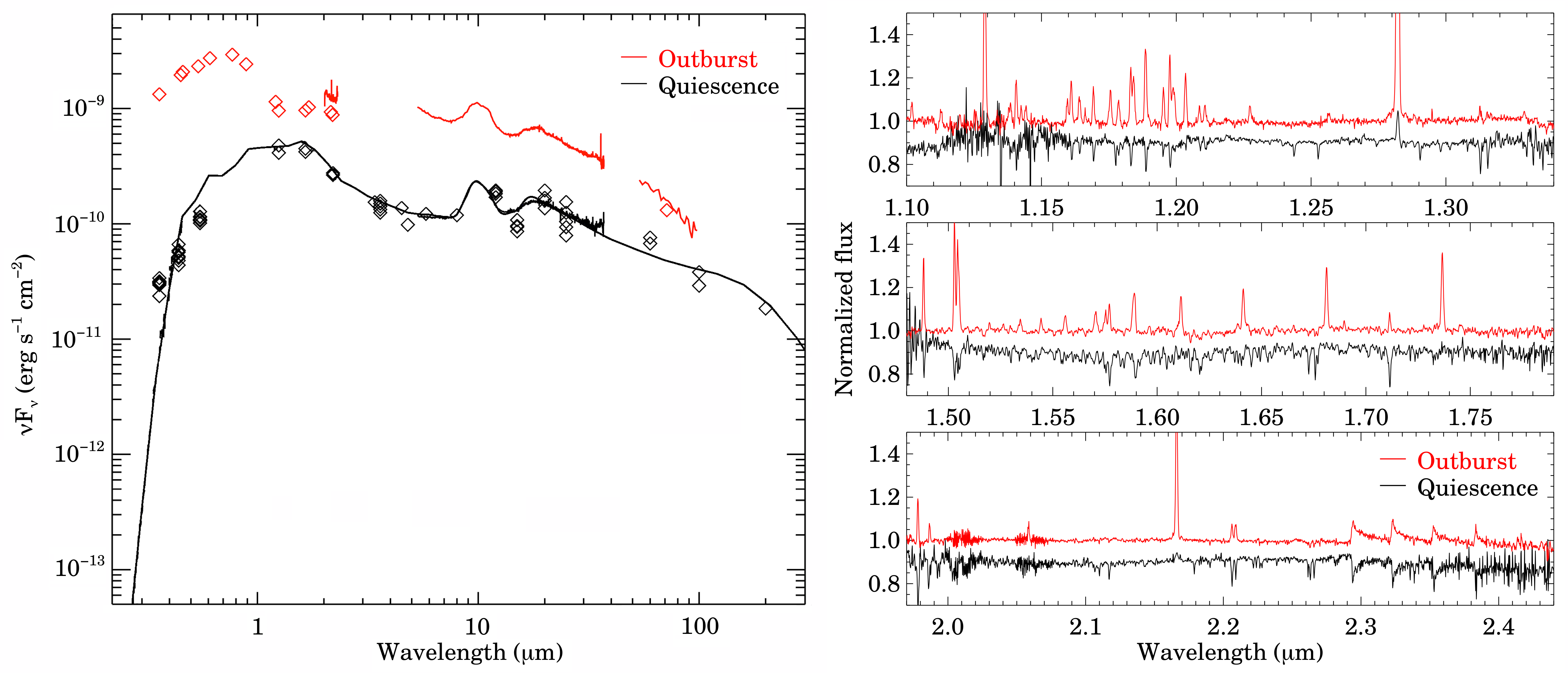}
\end{center}
\vspace*{-5mm}
\caption{EX Lup\index[obj]{EX Lup} itself, in quiescent and outburst states. The left panel shows the spectral energy distribution \citep{juhasz12}, while the right panel shows the (normalized and shifted) near-infrared spectrum \citep{rigliaco20}.\label{fig:beforeafter}}
\end{figure*}

Spectroscopically\index{spectroscopy}, in quiescence, metallic lines such as those of Na, Ca, K, Fe, Ti, and Si, as well as the CO bandhead feature are typically seen in absorption in optical and near-infrared spectra, and the stars appear like regular late-type young stellar photospheres. In outburst, when the disk increases in luminosity and can outshine the stellar photosphere, most absorption lines turn into emission, and many emission lines become stronger (\citealt{lorenzetti09, rigliaco20}; see also Figure~\ref{fig:beforeafter}).

EX~Lup type objects in outburst indicate ongoing magnetospheric accretion\index{magnetospheric accretion} from the disk onto the star. Therefore, not only the luminosity increase, but also their emission line fluxes (e.g., Paschen $\beta$ or Brackett $\gamma$) can be used to estimate accretion rates. The peak $\dot M_{\rm acc}$ is on the order of a few times $10^{-7}\,M_{\odot}\,$yr$^{-1}$ in outburst \citep[e.g.,][]{aspin10,juhasz12}. The relatively short timescale of the outbursts allows detailed studies of the differences between the high and low accretion states.

For example, \cite{sicilia-aguilar12,sicilia-aguilar15} used optical spectra of EX~Lup\index[obj]{EX Lup} in outburst and in quiescence to suggest very stable accretion channels, with similar structure but different throughput, at different times: $2{\times}10^{-7}\,M_{\odot}$yr$^{-1}$ in outburst and around $10^{-10}-10^{-9}\,M_{\odot}$yr$^{-1}$ in quiescence (\citealt{sipos09}; see also Figure~\ref{fig:beforeafter}). V1118~Ori\index[obj]{V1118 Ori} is another example of a repeating burster with different spectroscopic presentation in its low and high states \citep{herbig08,giannini20}. An increasingly large number of YSO bursts are being discovered with the spectroscopic characteristics of EX Lup type objects, with ESO H$\alpha$-99\index[obj]{ESO H$\alpha$-99} \citep{hodapp19} a recent example.

\textbf{V1647~Ori outbursts:} The discovery of the outburst of V1647~Ori\index[obj]{V1647 Ori} in 2004 revealed that some objects do not fit well into the two traditional categories of EX~Lup and FU~Ori type objects, even though these groups already encompass objects with a wide range of properties. V1647~Ori\index[obj]{V1647 Ori} has displayed multiple eruptions, with a duration longer than typical for EX~Lup objects but shorter than typical for FUors \citep{ninan13}. Some of the spectral characteristics of V1647~Ori\index[obj]{V1647 Ori} also resemble those of FUors, like water absorption, weak metals, and \ion{He}{1} absorption, while others are more characteristic of EX Lup type objects in the high state, including Pa $\beta$ and the CO overtone bandhead in emission \citep{briceno04,acosta-pulido07}. V1647~Ori\index[obj]{V1647 Ori} is also more deeply embedded than typical EX Lup type objects, and in this regard is similar to the younger FUors \citep{mosoni13}. 

Since the discovery of V1647~Ori, and especially over the past decade, a wide range of bursting and outbursting YSOs has been established. As amplitudes overlap between the outburst categories (Figure~\ref{fig:amp_vs_time}), and do not always go hand-in-hand with exhibited spectral properties, the prospects for definitive classification seem to be dimming rather than brightening. Some examples follow.

The V1647~Ori analogy can be made for objects that are spectroscopically quite similar to EX~Lup sources, but with much more slowly developing bursts, such as Gaia19bey\index[obj]{Gaia19bey} \citep{hodapp20}, or with longer duration outbursts than typical of EX Lup type objects, such as ASAS-13db\index[obj]{ASAS-13db} \citep{sa17}. The V1647~Ori category is, furthermore, a convenient place to park sources that are not quite as extreme in amplitude as FUors, yet have a more complicated spectral appearance than the emission-dominated spectra of outbursting EX~Lup stars. Gaia19ajj\index[obj]{Gaia19ajj} \citep{hillenbrand19b} is one such example with a large-amplitude rise and mixed spectrum.  Another interesting case is that of VVV-v721\index[obj]{VVV-v721} \citep{guo20}, which had a top-hat-like lightcurve similar to that of V1647 Ori (Figure \ref{fig:midir}) yet exhibited FU~Ori-like absorption features.

A similarly vexing source, in an even more embedded Class 0 phase, is HOPS 383\index[obj]{HOPS 383}, for which no spectrum could be obtained. Almost uniquely, this outburst was identified via its brightening by a factor of 35 at 24 \micron\ and a factor of 8 near 5 \micron\ \citep{safron15}. Follow-up at near-infrared \citep{fischer17b}, mid-infrared \citep[][see also Figure \ref{fig:midir}]{grosso20}, and sub-mm \citep{leeyh21} wavelengths suggested it had declined within several years. It is most likely V1647 Ori-like.

At present, the V1647 Ori category \citep{contreraspena17b} is not well defined, and the group is established perhaps more via what the objects are not, than by any unifying physics describing what they are. It is unclear if the grouping should include all objects with the lightcurve properties of V1647~Ori, regardless of their spectral presentation, or if it should contain only sources which have the exact mix between EX~Lup and FU~Ori properties in {\it both} lightcurves and outburst spectra that V1647~Ori itself does.

Both the EX~Lup and the V1647~Ori prototypes have repeating bursts/outbursts\index{outbursts}. A subset of the moderate-amplitude accretion brightenings that fall under these categories seem to feature not only repeated brightenings, but quasi-regular brightenings due to accretion changes. Examples include V371~Ser\index[obj]{V371 Ser} \citep{hodapp12, lee20}, V2492~Cyg\index[obj]{V2492 Cyg} \citep{covey11, hillenbrand13, kospal13}, and V347~Aur\index[obj]{V347 Aur} \citep{dahm20}. It is currently unclear if these sources are undergoing more cyclic rather than stochastic disk instabilities\index{disk instabilities} like EX~Lup and V1647~Ori exhibit, or if their quasi-periodic bursts relate to orbiting companions.

\textbf{FU~Ori outbursts:} One distinction between FU~Ori outbursts\index{outbursts} and the other flavors of accretion bursts described above is that the accretion rates are high enough to switch the accretion geometry from the typically dominant magnetospheric flow to one in which the accretion breaks through the magnetospheric barrier and reaches the star in roughly the equatorial plane \citep{hartmann16}. Bohdan Paczy\'nski in 1975 \citep[as noted by][]{trimble76} first suggested instabilities\index{disk instabilities} in a circumstellar accretion disk as a possible explanation for the outbursts of FU~Ori\index[obj]{FU Ori} and V1057~Cyg\index[obj]{V1057 Cyg}, the two FUors known at that time. As additional outbursting young stars were discovered in subsequent years (V1515~Cyg\index[obj]{V1515 Cyg}, \citealt{wenzel75}; V1735~Cyg\index[obj]{V1735 Cyg}, \citealt{elias78}; V346~Nor\index[obj]{V346 Nor}, \citealt{reipurth83}), more evidence accumulated in favor of the enhanced accretion disk model \citep{hartmann85}.

FUors in outburst have several common spectroscopic\index{spectroscopy} characteristics that are widely interpreted as forming in a viscously\index{viscous heating} heated accretion disk that dominates the spectrum. In the optical, these include absorption spectra similar to an F or G supergiant, broad blueshifted absorption lines indicating strong and variable winds, a P~Cygni (or pure absorption) profile for the H$\alpha$ line, and strong Li absorption at 6707 \AA{} \citep{hartmann96}. In the infrared, they resemble K or M giants or supergiants, with strong CO bandhead absorption, weak metal absorption, water and TiO or VO absorption bands, along with strong blueshifted \ion{He}{1} absorption \citep{connelley18}. High-resolution spectra show decreasing line width with increasing wavelength in some FUors (FU~Ori\index[obj]{FU Ori} and V1057~Cyg\index[obj]{V1057 Cyg} in \citealt{kenyon88}; V960~Mon\index[obj]{V960 Mon} in \citealt{park20}).

Both the linewidth behavior and wavelength-dependent spectral type can be understood in terms of a rapidly accreting disk that overshines the central protostar at all wavelengths. In this picture, the absorption lines form in a disk that is heated from the midplane and has a cooler ``atmosphere.'' Longer wavelengths trace colder, more distant, more slowly rotating disk regions, leading to a wavelength-dependent Keplerian broadening of the lines. For FU~Ori\index[obj]{FU Ori} itself, interferometric observations \citep{labdon21} have directly measured the expected $T(r)\propto r^{-3/4}$ profile expected from a disk.  

\looseness=1
FUor lightcurves show a large variety in the rise, peak duration, and decay. Considering the initial outbursts, among the classical examples, FU~Ori\index[obj]{FU Ori} and V1057 Cyg\index[obj]{V1057 Cyg} increased their brightness within about a year, while V1515~Cyg\index[obj]{V1515 Cyg} was brightening slowly for two decades \citep{hartmann96}. As additional FU~Ori stars beyond the first three have been identified, the rise times have been generally only poorly constrained. However, during the past decade, wide-field, moderate cadence, continuous photometric monitoring\index{monitoring} surveys have increased the number of identified FU~Ori-like events. The rise times have continued to exhibit dispersion, ranging from months (e.g., HBC~722\index[obj]{HBC 722}; \citealt{semkov10,miller11}) to years (e.g., V900~Mon\index[obj]{V900 Mon}; \citealt{reipurth12}).

\begin{figure}[!t]
\includegraphics[width=\columnwidth,angle=0]{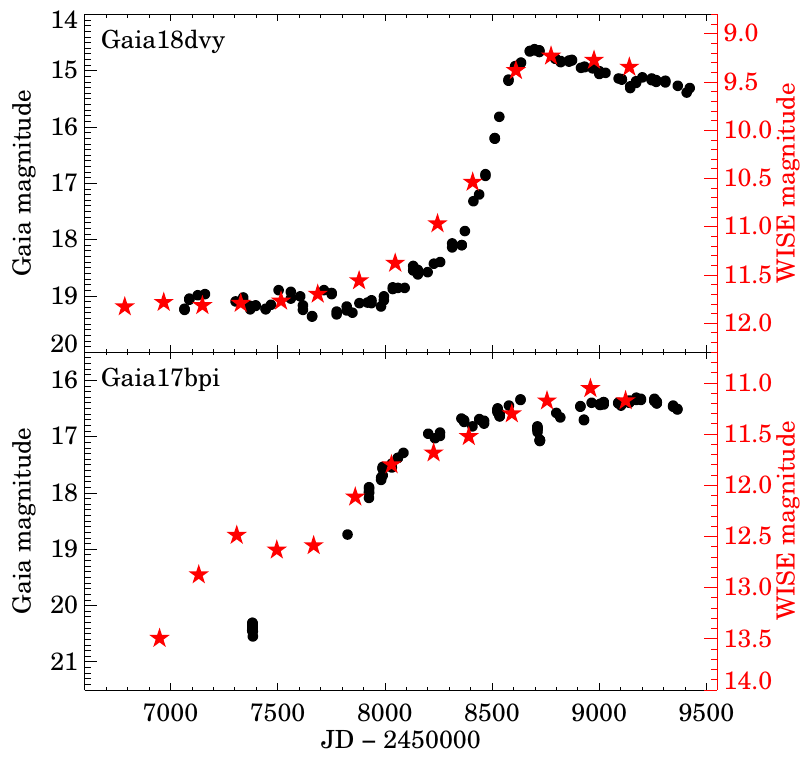}
\caption{Optical (Gaia; black) and infrared (WISE; red) lightcurves of Gaia18dvy\index[obj]{Gaia18dvy} \citep{szegedi-elek20} and Gaia17bpi\index[obj]{Gaia17bpi} \citep{hillenbrand18}.} \label{fig:gaia_wise}
\end{figure}

Intriguingly, the outbursting sources Gaia17bpi\index[obj]{Gaia17bpi} and Gaia18dvy\index[obj]{Gaia18dvy} have been suggested to have increased their brightness in the mid-infrared\index{mid-infrared} some months to years before their large-amplitude optical increases \citep[see Figure~\ref{fig:gaia_wise};][]{hillenbrand18,szegedi-elek20}. As infrared monitoring is a relatively recent development, the ubiquity, or lack thereof, of this phenomenon is currently unknown. One interpretation is that FU~Ori outbursts perhaps begin in a cooler region of the disk and propagate inward such that the hotter optical outburst is observed later.

While the lightcurve shapes show a large variety, FUor outbursts have in common a detected, sustained increase in bolometric luminosity and long decay timescale, perhaps exceeding typical human lifetimes. The luminosities, however, span a large range. Many of the FUors discovered in the twentieth century have luminosities of  several 100\,$L_\sun$, while other more recently discovered ones have luminosities of only 10--50\,$L_\sun$.

Under the assumption that the luminosity increase is purely due to accretion, and that the luminosity is dominated by the disk, the corresponding accretion rate $\dot M_{\rm acc} \approx L_{\rm acc}R_{*}/GM_{*}$ can be derived by matching the luminosity in fitting an SED, which requires understanding the source distance and extinction. Some FUors have peak accretion rates in excess of $10^{-4}\,M_{\odot}\,$yr$^{-1}$ (e.g., V582~Aur\index[obj]{V582 Aur}, \citealt{zsidi19}; V1057~Cyg\index[obj]{V1057 Cyg}, \citealt{szabo21}), while others only reach $10^{-6} - 10^{-5}\,M_{\odot}\,$yr$^{-1}$ (e.g., V2775~Ori\index[obj]{V2775 Ori}, \citealt{carattiogaratti11}; HBC~722\index[obj]{HBC 722}, \citealt{kospal16}; Gaia17bpi\index[obj]{Gaia17bpi}, \citealt{rodriguez21}). These numbers should be treated with caution as they depend on the (sometimes) poorly constrained stellar mass and disk geometry. Nevertheless, they probably reflect a real spread in FU Ori accretion rates. The strongest, longest outbursts may significantly and rapidly increase the mass of the protostar.

Regarding decay times, FU~Ori\index[obj]{FU Ori} has faded by only about 1\,mag since its outburst more than 80 years ago \citep{kenyon00}, while V1057~Cyg\index[obj]{V1057 Cyg} and V1515~Cyg\index[obj]{V1515 Cyg} have faded by 2--4\,mag in just a few decades \citep{szabo21,szabo22}. As they have faded, these objects have maintained their multi-temperature, disk-broadened spectral appearance. The fading is not always monotonic, as would be expected if disk cooling were the only physical effect.

The decay lightcurves for some FUors include short-timescale variability.   Flickering seen in precise space-based photometry has been interpreted as brightness asymmetries due to rotating inhomogeneities in the inner disk (FU~Ori\index[obj]{FU Ori}, \citealt{siwak18}; Bran~76\index[obj]{Bran 76}, \citealt{siwak20}; HOPS 383\index[obj]{HOPS 383}, \citealt{morales-calderon11}). Periodic variability\index{variability} is sometimes interpreted as a possible consequence of binarity (V1057~Cyg\index[obj]{V1057 Cyg}, \citealt{kopatskaya13}; V960~Mon\index[obj]{V960 Mon}, \citealt{hackstein15}). Sharp drops in brightness have been identified via multi-filter photometry to follow the interstellar reddening path, and hence interpreted as dust condensations \citep[e.g.,][for V1057 Cyg and V1515 Cyg]{clarke05}. 

We note that some objects that have been considered likely FUors early on during their outbursts -- albeit without exhibiting all confirming spectroscopic signatures of the class -- later undergo a significant decrease or halt in their accretion.  Recent examples include V899~Mon\index[obj]{V899 Mon} \citep{ninan15,park21b}, V346~Nor\index[obj]{V346 Nor} \citep{kraus16,kospal17}, V960 Mon\index[obj]{V960 Mon} \citep{park20}, and OO Ser\index[obj]{OO Ser} \citep{kospal07}, all of which faded monotonically back to their presumed quiescent brightness over about 5 years following their peaks.  These sources may be more appropriately considered as V1647 Ori type sources, even though they exhibited early signs of FU~Ori-like spectra.  Alternately, they may simply be short-lived FU~Ori type outbursts.  In these cases, the accretion fades carry information about the structure of the circumstellar matter and put constraints on how the accretion proceeds in these outbursting systems, which we can not yet access for the longer-lived FU~Ori outburst objects.

\textbf{The diversity of optical/infrared outbursts:} Despite the wide range of behaviors already described above, several recently identified outbursting objects have exhibited even more unusual rise times, amplitudes, or spectra.

\cite{herczeg16} presented the case of an object with an FU~Ori-like absorption and wind spectrum, but which rose to peak brightness in $<$24 hours and decayed on a few-month timescale. In the case of WISE 1422$-$6115\index[obj]{WISE 1422$-$6115} \citep{lucas20}, the outburst amplitude is an ``exceptional'' 8 mag in the near- and mid-infrared.  This source is not optically visible, with a $K$-band spectrum that is completely featureless, lacking either emission or absorption. For PTF~14jg\index[obj]{PTF 14jg} \citep{hillenbrand19a}, the relatively short rise and large-amplitude event showed an absorption spectrum that is atypically hot for an outbursting YSO and a featureless spectrum in the $K$ band. The latter two objects faded on timescales of 5--10 years. These examples illustrate the widening range of outburst phenomenology, {labeled {\it peculiar} by \citet{connelley18}}, to be explained beyond the already diverse EX~Lup, V1647~Ori, and FU~Ori classes. On the other hand, these three specific outstanding sources may not be YSOs at all, despite their apparent association with star-forming regions.

As implied throughout our discussion, there is significant diversity in lightcurve shapes and spectroscopic presentation of YSO outbursts. Precious few sources seem to fit the classical definition of an FU~Ori lightcurve {\it and} spectrum. More examples are known of the repeating EX~Lup type events. The majority of sources recently discovered in outburst appear to have ``mixed'' characteristics, exhibiting absorption and emission features in their spectra that are not entirely aligned with the expectations. For example, some objects with disk-like absorption spectra in the blue optical do not seem purely FU~Ori-like, since they also show emission line features at redder wavelengths. The intermediate V1647~Ori category is a convenient bin for such hybrids, though it may be the case in the future that further sub-classification efforts will be possible, once a larger sample of well-characterized quiescence-to-peak-to-quiescence outbursts is available.

Another axis of diversity is in the multiwavelength amplitudes of the accretion bursts and outbursts. EX~Lup type objects have long been appreciated to have blue outbursts, with larger amplitudes at shorter wavelengths \citep[e.g.,][]{giannini20}. This color change is to be expected in the scenario of moderate enhancements in the magnetospheric accretion\index{magnetospheric accretion} rate, with more mass infalling at similar velocities, resulting in higher temperatures and hence bluer optical colors. For the FU~Ori outbursts, the expectation is of a globally enhanced accretion rate,   involving more of the disk, and resulting in an increased bolometric luminosity across a broader range of wavelengths. There is limited empirical color information in the pre-outburst states, though with evidence for colors becoming slightly bluer (few tenths of a magnitude), at least temporarily, during the outbursts near peak \citep{szegedi-elek20}.

Finally, the well-studied bursts and outbursts discussed above have systematically larger amplitudes, measured in the optical, than do those bursting sources identified through mid-infrared variability (discussed next). In the scenario where $\dot M_{\rm acc}$ is temporarily enhanced only in the inner few tenths of an au of the disk, the limited area of the hot accretion zone may naturally explain this amplitude discrepancy. We note the importance of more detailed theoretical modeling showing what the multiwavelength lightcurves of accretion outbursts should look like, in order to constrain the outburst mechanisms.

\subsubsection{Variable Accretion in Protostars Revealed via Long-Wavelength Monitoring}
\label{sec:midIR}

\begin{figure*}[!t]
\begin{center}
\includegraphics[width=\textwidth]{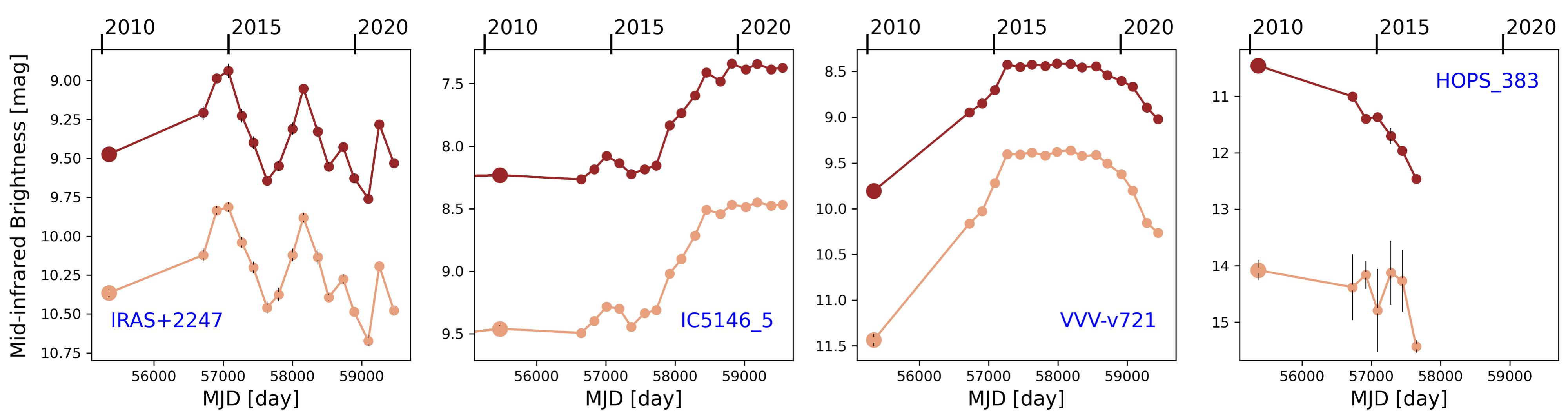}
\end{center}
\vspace*{-5mm}
\caption{Examples of YSO variability in the mid-infrared (orange = W1 mag; maroon = W2 mag) monitored by WISE/NEOWISE. IRAS 04302+2247\index[obj]{IRAS 04302+2247} shows stochasticity, or potentially years-long quasi-periodicity; IC5146\,5 (2MASS J21533472+4720439\index[obj]{2MASS J21533472+4720439}) appears to be in the middle of a burst; VVV-v721\index[obj]{VVV-v721} (2MASS J16394877$-$4548480) and HOPS~383\index[obj]{HOPS 383} burst then faded within a few years, and likely belong in the V1647~Ori category (see \S~\ref{s.gulps}).}\label{fig:midir}
\end{figure*}

The historical description of accretion outbursts developed mostly through optical monitoring and follow-up spectroscopy. However, during the main stages of stellar growth, protostars\index{protostars} are embedded in an optically thick envelope and located deep within their parent cloud. Their growth and any accretion variability are detectable primarily at infrared and radio wavelengths, with observations that pierce through the opaque dust.

While variability\index{variability} at long wavelengths has been established to be common, the interpretation and generalization are still challenging. Several surveys at longer wavelengths have been designed to statistically\index{statistics} evaluate changes in accretion rate. \citet{offner11} estimated that 25\% of a star's main-sequence mass could be accumulated in protostellar outbursts, and \citet{fischer19} found a similar fraction by considering the estimated durations, amplitudes, and duty cycles of protostellar outbursts in Orion. Mid-IR monitoring\index{monitoring} (see Figure \ref{fig:midir} for examples) and submillimeter surveys for bursts are still in their infancy compared to optical studies, but they have the potential to give a systematic understanding of mass assembly at early times.

The VVV Survey $K$-band monitoring has revealed a large number of candidate outbursting protostars, primarily in the galactic plane \citep{contreraspena17a,contreraspena17b,guo20}. Such studies require additional spectroscopic confirmation to discriminate between young stars and the many contaminants in the monitoring regions. The outbursting YSOs, a much less biased sample than the historical set of outbursts, demonstrate the diversity in outburst duration and amplitude. Most diverge from the FU~Ori/EX~Lup classification system, with some resembling the intermediate duration of V1647 Ori. Reaching conclusions as to the overall outburst frequency and variability level by evolutionary stage remains challenging, because the parent population is unknown.

Mid-IR surveys, particularly the all-sky monitoring of NEOWISE, have enabled outburst searches on carefully selected samples of protostars. Mid-IR searches for large changes were initially carried out by comparing two epochs of Spitzer\index{Spitzer} and/or WISE\index{WISE} observations, spanning up to six years \citep{scholz13,rebull14,fischer19}. More robust lightcurve characterization and statistics\index{statistics} were enabled by including $>$10 data points spanning $>$5--7\,yr \citep{lucas20,park21,zakri22}. Among the identified protostars in \cite{park21},  $\sim$35\% show continuous stochasticity at the 6 month cadence without a discernible timescale and another 20\% have years or longer apparently secular events with moderate, factor of two changes in brightness; only a few percent show identifiable discrete short-timescale ($<$1\,yr) events. While much of the observed mid-IR variability has similar characteristics to the optical/near-IR pre-main sequence routine accretion variability (\S~\ref{sec:obs}), for protostars there appears to be a stronger long-timescale, low-amplitude component, likely associated with disk accretion physics.

The mid-IR studies commonly set a threshold of 1 or 2 mag to distinguish departures from routine variability. This is consistent with our definition of ``bursts'' but not ``outbursts'' in Table~\ref{tab:nomenclature},  which is built on optical observation.  In the infrared, flux differences between FUor objects and typical protoplanetary disks are lower than in the optical \citep[see][for illustration]{hillenbrand22}. At least two stars with only a $\sim$2\,mag brightening in the 3--5\,\micron\ range  (which would correspond to $\sim$4\,mag in the near-infrared) show the near-IR spectral characteristics of FU~Ori stars (V2775~Ori\index[obj]{V2775 Ori}, \citealt{fischer12} and Gaia18dvy\index[obj]{Gaia18dvy}, \citealt{szegedi-elek20}).

Typical burst/outburst intervals for embedded protostars have been estimated to be $\sim$1000\,yr from investigating the objects that changed by $\sim$2\,mag across {Spitzer}\index{Spitzer} and {WISE}\index{WISE} epochs separated by 6.5\,yr \citep{fischer19} and sustained over at least several epochs in 7 years of NEOWISE monitoring \citep{park21}. On the other hand, \citet{rebull14} did not note any large bursts in a large sample when comparing between epochs separated by 6--7 years. Findings depend upon classification choices, including whether to interpret a fading star as a candidate \citep[as implemented by][]{park21} and what to call a burst or an outburst. The frequency of once per $\sim$ 1000\,yr appears to be much higher than the frequency in older disks \citep{hillenbrand15,contreraspena19}, but comparisons are not consistent across studies, with large uncertainties in both the types of objects being detected and in the parent samples. Additionally, candidates identified from lightcurves require subsequent confirmation.

An important caveat to the outburst frequency is that understanding the nature and cause of the burst requires a spectrum, which is difficult or impossible for deeply embedded sources but feasible for flat-spectrum or late Class~I protostars. At mid-IR wavelengths, the radiative transfer can be complicated, since the measured emission may vary due to geometric changes in the inner disk, and the change in disk emission may not directly correlate with a change in accretion rate. Infrared emission may also escape through outflow cavities, where the envelope opacity is lower, providing opportunities to identify variability in reflected light \citep[e.g.,][]{muzerolle13,yoon22} but introducing challenges in interpretation.

At submillimeter\index{submillimeter} wavelengths, variability is cleanly interpreted as a temperature change in the dust following the variable protostar luminosity, since radiative transfer complications should be negligible when measuring volumes \citep{johnstone13,baek20}. Studies that combine sub-mm and IR observations confirm the close correlation between sub-mm and protostellar emission for individual objects \citep{safron15,lee20,szabo21} and across small samples \citep{contreraspena20}.

Monthly cadence monitoring in the sub-mm reveals long-term, many-year, secular variability trends on $\gtrsim$20\% of sources at the $\sim$10--30\% level \citep{johnstone18,leeyh21}. The sub-mm brightness changes primarily reflect a change in the equilibrium temperature of the dust in the enshrouding envelope, which responds to source luminosity changes only weakly, with $T_d \propto L^{\sim1/5}$ \citep{johnstone13, macfarlane19a}. Thus, the sub-mm brightness variations indicate significantly larger fractional changes in protostellar luminosity of 1.5--3 as demonstrated for a sample of sub-mm variables through comparison with mid-IR variability \citep{contreraspena20}.

\looseness=1
The statistics on long-term, many-year, sub-mm variations are similar to those found by the mid-IR surveys discussed above; however, the sub-mm observations fail to uncover a class of continuously stochastic sources, either due to the poor sub-mm measurement sensitivity  or possibly a smoothing out of the lightcurve by the finite, month-long equilibration time of the protostellar envelope \citep{johnstone13}. These surveys have only recently become possible with techniques based on relative brightness measurements and are still limited to small numbers of protostars \citep{mairs17b}. As a consequence, large outbursts have not yet been discovered at sub-mm wavelengths. Nevertheless, for a few sources the sub-mm lightcurves are allowing for physical interpretation. Several bursting systems previously identified at shorter wavelengths show correlated long-term sub-mm behaviour \citep[e.g., V1647~Ori, HOPS~383, V371~Ser;][]{leeyh21}, with V371 Ser\index[obj]{V371 Ser} (EC~53) showing an $\sim$18 month sub-mm periodicity \citep{yoo17, lee20} that is also seen in the near-IR {as discussed in \S~\ref{s.gulps}}.

Long-wavelength observations are also beginning to uncover brightness variations in high-mass star-forming regions\index{massive star formation}. As massive star evolution proceeds more quickly -- at all evolutionary stages -- than the development of low-mass stars, the protostellar and pre-main sequence stages for $M> 5\,M_\odot$ remain entirely heavily embedded. The known outbursts in this mass range are rare. Examples include the variable high-mass protostar NGC6334I-MM1\index[obj]{NGC6334I-MM1}, which brightened in the mm, but is challenging to study since it is both saturated in mid-IR imaging from WISE/NEOWISE and invisible in the red-optical to 22 mag \citep{hunter17}. Other outbursts are found in S255\index[obj]{S255} \citep{caratti-o-garatti17} and M17\index[obj]{M17} \citep{chen21}. For high-mass protostars, a potential indirect indicator of accretion variability is Class II methanol maser flaring \citep[e.g.,][]{macleod18}. G358.93$-$0.03-MM1\index[obj]{G358.93-0.03-MM1} was initially noticed due to such a maser burst \citep{sugiyama19} and was later confirmed to have undergone a far-IR brightness increase by a factor of a few \citep{stecklum21}. For this source, the wave of radiation generated methanol maser emission from the disk, which was used to measure a precise protostellar mass \citep{burns20}. Several candidate bursts have also been identified on massive protostars with NEOWISE monitoring \citep{uchiyama19}. \cite{magakian19} even identified an optical outburst of a Class I source inferred to be a B star.

\subsection{\textbf{Indirect Signatures of Variable Accretion}}

\subsubsection{Spectra Indicative of Viscous Accretion Disks}

FU~Ori outbursts persist for decades, perhaps even centuries. However, the brightness increase associated with any particular outburst may have been missed. Nevertheless, stars in an FU~Ori outburst state should show the unique spectral absorption signatures that arise from a disk photosphere, in particular the temperature dependence with wavelength, the line profiles consistent with disk rotation, as well as a disk-like ultraviolet, optical, and near-infrared spectral energy distribution (see \S~\ref{s.gulps} for more details).

L1551~IRS~5\index[obj]{L1551 IRS 5} and Bran~76 (V646~Pup\index[obj]{V646 Pup}) are examples of YSOs that share spectral characteristics with FUors, but no known outburst has been observed \citep{harris80,reipurth02}. Such sources are known as FUor-like objects \citep{aspin03,greene08} and must be associated with strong disk accretion, likely in temporary outburst. Often such objects are faint and red, with only low-resolution spectra that render them difficult to distinguish them from late-type stars \citep{connelley18}. Only $\sim$10--15 FUor-like objects have been identified.

\subsubsection{Chemical Tracers of Outbursts}\label{sec:chem}\index{chemical tracers}

When a protostar undergoes a large accretion burst, the increased luminosity will heat the disk and envelope above the quiescent temperature, leaving a temporary or lasting mark on the chemical structure \citep[see review by][]{jorgensen20}. Some molecules will be dissociated and ices destroyed in the outburst, with recombination and reformation timescales that depend on the specific species and local density. The chemistry in the envelope and disk offers a historical record of outbursts in the most recent few millennia.

For envelopes, \citet{visser15} found that for a factor of 100 increase in luminosity, CO evaporates throughout the envelope, leading to an increased abundance of HCO$^+$, decreased abundance of N$_2$H$^+$, and a water snowline expanded by a factor of 10. The warmer envelope also leads to the evaporation of complex organic molecules \citep{taquet16}. These models provide a framework for interpreting measurements of CO snowlines. These are often located beyond the distance of the snowline expected from the present-day luminosity, as expected for recent bursts \citep[e.g.,][]{lee07,jorgensen15,frimann17}. \citet{hsieh19} leveraged the differential reformation timescales of 1,000 years for H$_2$O ice and 10,000 years for CO ice to infer past luminosities of 10--100 \lsun\ for a sample of Class 0/I sources in Perseus. These luminosities are much higher than present-day values and, if interpreted as bursts, these events occur once every 2400\,yr for Class 0 objects and once every 8000\,yr for Class I objects. The Class~I protostellar binary L1551~IRS~5\index[obj]{L1551 IRS 5} is a possible example of a disk/envelope structure with complex chemistry, including complex organic molecules, driven by an FUor-like outburst of the northern component \citep{bianchi20}.

For protoplanetary disks, the chemical effects are also significant but do not persist for as long as in the envelope. Timescales are shorter in the dense inner disk and longer in the outer disk. \citet{rab17} found that CO sublimates during the outburst, then once the YSO is back in quiescence the CO freezes out in an inside-out manner that produces distinct observational signatures, such as a two-component Gaussian intensity profile post-outburst. For a 200\,$L_{\odot}$ outburst luminosity lasting 50 years, \citet{molyarova18} found that gas-phase abundances of molecules significantly increased due to the outburst, mainly because of the sublimation of their ices. The abundances of CO, NH$_3$, C$_2$H$_6$, and C$_3$H$_4$ return nearly instantaneously, while H$_2$CO and NH$_2$OH remain in the gas phase for 10--10$^3$\,yr after the end of the outburst.

Despite significant uncertainties related to interpretation of chemical diagnostics, the prediction that complex organic molecules will evaporate off of grains at high temperatures during outbursts has been confirmed for the inner disk of the outbursting star V883 Ori\index[obj]{V883 Ori} \citep[e.g.,][]{hoff18,lee19}. These indirect probes provide a powerful diagnostic of the temperature history of protostars and can be used to identify stars that experienced outbursts at relatively recent times \citep{anderl20}.

\subsubsection{Outburst Signatures in Outflows}\label{sec:outflow}\index{outflows}

Outflows driven by young stars can be used as an indirect tracer of accretion variability. Since the mass ejection originates from disk accretion, any variability in the accretion process should lead to corresponding variability in the ejection process. This can manifest as clumpy structures in the jets and outflows, where individual clumps may plausibly be tied to different accretion/ejection events, with the clump spacings and kinematics together used to estimate the duration between individual accretion (and ejection) bursts. While this idea has been around for at least three decades \citep{reipurth89} and was reviewed in the previous Protostars \& Planets VI \citep{dunham14,frank14}, improved facilities have led to a proliferation of outflow characterization in recent years, including a known FUor \citep{kospal08}.

Typical durations between bursts inferred from outflow signatures of variability are a few hundred to a few thousand years \citep[e.g.,][]{plunkett15,zhang19,nony20,vazzano21}. 
There are substantial uncertainties inherent in this type of analysis, including projection effects on both the clump spacings and observed velocities, assumptions related to whether detected clumps are locally entrained or ejected from the driving source, and assumptions related to the history of the velocity of the outflowing material. Furthermore, \citet{vorobyov18} showed in models that longer burst durations could also be present but undetected in outflow clumps due to the large distances that such clumps would travel over such long durations.

While it is now clear that some outflows exhibit signatures of variability in ejection properties, the connection to accretion events is reasonable but mostly speculative, with only a few tentative connections that have been identified \citep[e.g.,][]{ellerbroek14,garufi19}.  Wind absorption features are one of the defining characteristics of FUor outbursts \citep{herbig03,sicilia-aguilar20,szabo21}, including enhanced emission from a bipolar wind associated with an outburst of Z CMa\index[obj]{Z CMa} \citep{benisty10}. However, it is unclear how or even whether the wind from the inner disk translates to enhanced molecular outflows. Significant progress has been made in linking maser emission from outflows to outbursts for high-mass stars \citep{burns16}, with the prospect of developing a scaling relationship, but these may not be applicable to low-mass stars or to bursts in the distant past.

Outflows offer a rich historical archive of accretion history over timescales of thousands of years. However, to exploit this archive, large uniform surveys are needed to firmly establish and quantify the relationship between outflow variability and accretion variability.

\section{\textbf{UNRESOLVED ISSUES}}\label{sec:unresolved}

\subsection{\textbf{What Drives Accretion Variability?}}\label{sec:physics}

Directly observed and indirectly inferred evidence argue for unstable mass accretion through the disk during protostellar assembly. In contrast, short-duration events are likely driven by instabilities\index{disk instabilities} in the flow from the very inner disk onto the forming star, with the observed variability probing the mechanisms responsible for this mass transfer (\S~\ref{sec:obs}). Intermediate between these two extremes, the observed moderate-amplitude years-to-decade-long secular variability apparently common in protostars (\S~\ref{sec:midIR}) probes inner-disk viscous accretion times and thus may yield useful constraints on the physical processes mediating inner disk accretion.

Whether these accretion variations are located along a continuum of instabilities, as perhaps indicated by the intermediate events, or whether the large and small bursts are distinct phenomena, remains unclear.  The present observational classification schemes are insufficient and the statistics are too poor to precisely populate the amplitude-timescale diagram (see Figure \ref{fig:amp_vs_time}). While this question is still ill-formed observationally, one can instead describe the problem in terms of the physics that governs the outbursts of different sizes and amplitudes. To consider that question, in this section we describe the physics that may trigger and end outbursts. We also look at systematic differences between FU Ori disks and other YSO accretion disks.

\subsubsection{Potential Accretion Variability Mechanisms}

Fundamentally, the observed timescales associated with accretion variability, whether episodic, periodic, or stochastic,  depend on the physical conditions within the disk and the processes responsible for producing the unstable flows. Furthermore, the amplitude of the variability depends on the ability of the process to extract energy and angular momentum from the disk. For events at disk radius $R$ and modulated by gravity, the local dynamical time\index{dynamical timescales},
\begin{equation}
t_{\rm dyn}(R) = 1\,\left(\frac{M_*}{M_\odot}\right)^{-1/2}\,\left(\frac{R}{1\,{\rm au}}\right)^{3/2} {\rm yr},
\end{equation}
ranges from mere days, near the inner disk edge at several $R_*$, to centuries, for material located in the outer disk. On the other hand, if diffusive evolution of the disk material is important, as for example implied by the viscous $\alpha$-disk prescription for mass transport within the disk, then the diffusion time,
\begin{equation}
t_{\rm dif}(R) \sim 10^4\,\left(\frac{0.01}{\alpha}\right) \left(\frac{0.1}{H/R}\right)^{2} t_{\rm dyn}(R),
\end{equation}
sets the timescale, with $H$ the disk scale height, the typical steady-state $\alpha < 0.01$, and $H/R < 0.1$. Thus, even in the very inner disk the expectation is that viscous diffusion requires a minimum timescale of years. Furthermore, a thermal timescale\index{thermal timescales}, $t_{\rm therm}(R) \sim \alpha^{-1}\,t_{\rm dyn}(R)$, lies intermediate to the dynamical and diffusion times;
\begin{equation}
t_{\rm therm}(R) \sim 10^2\,\left(\frac{0.01}{\alpha}\right)t_{\rm dyn}(R),
\end{equation}
corresponding to months near the inner disk edge. Finally, at intermediate distances, disk accretion may be driven by magnetocentrifugal disk winds\index{disk winds} \citep[][see also chapters by Manara et al.\ and Pascucci et al.\ in this volume]{bai17,lesur21}. Radial scaling arguments for the evolution of episodic events associated with such accretion do not yet exist; however, the relevant event timescales are likely to be intermediate between $t_{\rm dyn}$ and $t_{\rm dif}$.

From the above argument, days-to-weeks timescales for observed variability are driven by the very inner disk, the disk-star interaction, or the accreting star. On these timescales we also observe non-accretion-driven brightness changes by stellar rotation (star spots) or chromospheric activity, and potentially inner disk warping capable of producing short-timescale variations in optical depth. Variable accretion may be induced by instabilities in the accretion columns connecting the star with the disk, but unless a feedback mechanism is invoked, the observed time-averaged mass accretion onto the star should be determined by the steady-state flow through the inner disk. Possible feedback mechanisms at the star-disk interaction region include magnetic gating of the disk mass flow, leading to quasi-periodic accretion \citep[e.g.,][]{dangelo12}, and reversal of the stellar magnetic field driven by dynamos \citep{armitage16}. On somewhat larger scales, convective or thermal instabilities\index{disk instabilities} within the inner disk, driven by and coupled to the varying accretion luminosity, can produce periodic or episodic mass flow onto the star \citep[e.g.,][]{kadam19, kadam20, pavlyuchenkov20}.

Longer-timescale accretion variability is expected to be driven by the mismatch between steady-state mass flow through the disk and local regions where the flow becomes stalled. For example, the magneto-rotational instability (MRI)\index{disk instabilities} is a favored mechanism for interchange of energy and angular momentum across inner disk material, leading to diffusive spreading of the disk but requiring a moderate level of ionization to couple to the disk magnetic field. The star provides ionizing radiation to the very inner disk \citep[e.g.,][]{flock17}, and cosmic rays may work at larger distances \citep[e.g.,][]{bae14}, but in between there is likely to be a region incapable of efficient MRI-induced transport \citep[e.g.,][]{zhu10}. This bottleneck, at radii where the accretion rate may be less efficient, disconnects the inner disk from a large reservoir of mass in the outer disk, and thus inner-disk viscous\index{viscous evolution} spreading dumps an ever decreasing amount of material onto the star, while the mass in the outer disk grows, spreads, and piles ever more material at the outer edge of this blockage. Potential solutions to this flow stagnation have been proposed in the literature, all of which lead to episodic mass accretion events -- in particular, heating-induced ionization within the blockage due to gravitational instabilities\index{disk instabilities} near its outer edge lead to episodic runaway events that efficiently transport material into the inner disk \citep[e.g.,][]{armitage01, martin11}.
Indeed, for FU Ori outbursts, the observed disk temperatures are high enough ($> 1000$\,K) for thermal ionization to activate MRI \citep{gammie96, desch15} over a large area of the inner disk, allowing for efficient accretion.

Alternatively, outer-disk gravitational instability can either induce fragmentation into clumps, which may then scatter on dynamical timescales into the inner disk and potentially be tidally disrupted and viscously diffuse onto the star \citep[e.g.,][]{vorobyov21}, or it can induce inner-disk MRI and large accretion events \citep{zhu09,bae14}. At the earliest stages of protostellar evolution, accretion from the envelope onto the disk should also lead to disk instabilities\index{disk instabilities} and perhaps  non-steady accretion \citep{bae15,lesur15}.

Observations of bursting and outbursting pre-main sequence stars provide time-dependent evidence of the changing conditions within the disk, including measurements of the disk radial temperature distribution and sequencing of the heating within the disk -- inner to outer versus outer to inner -- important constraints for the theoretical models. Likewise, statistical consideration of both the bursts/outbursts and the lower-amplitude routine accretion variability observed on inner-disk viscous timescales across all YSOs, protostellar and pre-main sequence, should restrict the range of possible accretion disk models.

\subsubsection{Means of Ending Outbursts}

Once an accretion burst or outburst has begun, its duration depends on the ability of the disk to continue supplying material at an enhanced rate. Mechanisms that rely on inner disk instabilities\index{disk instabilities} therefore have a shorter potential lifetime before exhaustion of the available disk material. Alternatively, for instabilities driven further out in the disk, both the timescales and mass reservoirs will be significant. The theoretical models discussed above primarily produce episodic accretion events due to an imposed mismatch in the accretion onto the disk and the steady-state throughput of the disk. In such cases, after each burst is complete, there will be an extended lull as the disk refills and returns to the unstable state. For example, \citet{banzatti15} found weaker molecular emission from EX Lup\index[obj]{EX Lup} after the burst compared to an in-burst spectrum, possibly the consequence of a decrease in molecular column density, i.e., the inner disk rapidly drained of material.  During the last stage of stellar assembly, however, it is possible that a disk instability reduces the remaining disk mass significantly in a single, last, gulp.

Some targets exhibit brief periods of quiescence in the middle of their outbursts. V1647 Ori\index[obj]{V1647 Ori} and V899 Mon\index[obj]{V899 Mon} showed breaks of a few months to a year \citep{ninan15}. In both cases, the second outburst was much longer than the first outburst and the break. The decay timescale of the first outburst was also significantly faster than the decay timescale of the second outburst. This may indicate that additional complexities are at work in ending some outbursts.

There is thus a rich potential for identifying outburst mechanisms through a quantitative understanding of their amplitudes, lifetimes, and, for those that repeat, their cadences. Complicating matters for observers, as noted at the start of this section, the viscous timescales within the disk are long, even at the very inner disk edge. Furthermore, the majority of the theoretical models show significant non-linear disk behavior mixed in with each burst. Thus, the triggering of an instability leads to a diverse string of accretion events, with a correspondingly broad spectrum of individual amplitudes and timescales.

It is therefore unlikely that any one observed accretion outburst system can be unambiguously associated with a specific underlying instability mechanism from its lightcurve alone. This degeneracy may be weakened through the aid of key observational diagnostics measuring the physical state of the protostar and disk.

\subsubsection{Why Are FU~Ori Disks More Massive and Smaller Than Other YSO Accretion Disks?}

\index{disk mass}
\index{disk radius}
In the enhanced disk accretion model, an important aspect is that there should be a mechanism that stops the material from falling onto the star continuously, leading to an accumulation of material in the inner disk; and also a mechanism to quickly increase the low, quiescent accretion by orders of magnitude, leading to an eventual depletion of the inner disk, which may be replenished later. Until recently, very few observational constraints existed for the spatial and velocity structure of the circumstellar disks and envelopes of young eruptive stars. Millimeter interferometers provide this crucial information. Using ALMA observations of the millimeter lines of various CO isotopologues, \citet{kospal17b} found that the infall rate from the envelope onto the disk in V346~Nor\index[obj]{V346 Nor} is a factor of a few higher than the quiescent accretion rate from the disk onto the star, hinting at a mismatch between the infall and accretion rates that models indicate can create a mass buildup and eventually cause an eruption \citep{zhu10}.

Millimeter continuum observations of young eruptive stars with ALMA and SMA revealed that FU~Ori disks are significantly more massive than EX Lup type disks, and that they are more compact than what is typical for normal T~Tauri disks \citep{cieza18, liu18}. Radiative transfer modeling of ALMA 1.3\,mm continuum observations of a sample of FUors and FUor-like objects revealed that disk masses are significantly larger than what was obtained from the measured millimeter fluxes assuming optically thin emission, suggesting that FUor disks are at least partially optically thick at this wavelength and suggesting that substantial amounts of mass may be hidden in their disks \citep{kospal21}. When compared with samples of regular Class~II and Class~I sources modeled in the same way, FUor disks were found to be systematically more massive and smaller in size. A significant fraction of FUor disks may even be gravitationally unstable. The small, massive disks imply significant mass reservoirs at small radii, i.e., in regions with small enough viscous timescales, which may provide an explanation for the eruptive nature of these objects.

That FUor disks appear distinctly more massive and smaller than typical T Tauri disks raises intriguing questions. Is it possible that there is a phase during the circumstellar disk evolution when the disk structure deviates from the typical one so that it can produce outbursts? Is it possible that FUor eruptions happen in normal disks, but the outburst itself modifies the disk structure? Is it possible that some disks are born with smaller radius and larger mass than usual, and later they follow a particular evolutionary path that leads to episodic eruptions? To answer these questions and determine if {\it all} YSOs go through FUor eruptions will require a well-understood framework for the initial distribution and evolution of gas and dust in disks from formation to the Class II stage.  

\subsection{Why Is the Rate of True FU~Ori Outbursts Still Poorly Constrained?}

Assuming that all YSOs undergo FU~Ori outbursts, attempts from optical discoveries place the FU~Ori outburst rate at one per $\sim 10^{5}$ yr \citep[e.g.,][]{contreraspena19}, while evaluations from the mid-IR yield outburst rates of one per $\sim 10^3$ yr \citep{fischer19,park21}. These numbers seem to confirm our expectations, from simple arguments and from theory, that protostars have a higher frequency of outbursts in the primary growth stages, when the disk is still being fed by the envelope. Despite this clean story, the rate of true FU~Ori outbursts remains highly uncertain. The rate requires an accurate number of events divided by the total number of young stellar objects that could have been detected to burst, with sufficient statistics\index{statistics} to be confident in the result \citep{hillenbrand15}. One further needs to consider evolutionary stage if we want to measure a time dependence. 

The numerator requires a sufficient number of outbursts for a statistically meaningful result and an accurate classification of those outbursts. True FUor outbursts are exceedingly rare and must be distinguished from other, far more frequent large-amplitude variability in the galactic plane (CVs, symbiotic stars, AGB pulsations) as well as from other flavors of large-amplitude but non-outburst variability that characterizes some young stars.  \cite{contreraspena19} found six among a large sample of Class II stars observed in epochs separated by 55 yrs, distinguishing them from 133 other young stellar objects with large changes, and 15,000 (possible) Class II stars that did not show such large changes. Studies over substantially longer timescales with increased samples are not obviously feasible.

\looseness=1
The numerator is even more challenging to measure for protostars. The threshold of brightness change to identify candidate FUor outbursts is typically $\sim 2$ mag in the mid-IR, compared with 3--5 mag in optical surveys. Once a candidate is identified, the historical lightcurve for context is often not available, and spectroscopic confirmation of a viscous disk may be impossible. Stringent criteria that may be applied for optically bright objects often cannot be applied to faint protostars, leading to looser classifications. The methodological differences and the desire to label an object as an outburst potentially lead to much higher measured rates of outbursts for protostars.

The denominators are also plagued with uncertainty. First, the parent population should be bona fide young stars with disks, which are now possible to define for optically-bright members within $1$ kpc using {Gaia} and {WISE} (a relatively small population) but challenging for infrared selections.  For any time dependence, protostars in the galactic plane remain difficult to distinguish from background AGB stars, {despite diagnostics that include emission from dense molecular gas \citep{heiderman15}, astrometry \citep[e.g.,][]{herczeg19}, and lightcurve shapes and analysis of maser emission \citep{leejl21}.} Second, since FUors are $\sim$5 mag brighter, they are detectable to a much larger volume than their quiescent counterparts; thus, volume-limited samples severely reduce the statistical significance in the numerator.

\subsection{Bridging Theoretical Models and Observational Constraints}

The theoretical models and numerical simulations reveal an impressively rich parameter space of physical processes that can produce episodic mass flows onto forming stars \citep[e.g.,][]{vorobyov21,pavlyuchenkov20,maksimova20}. However, it is very difficult to compare them directly with observations. This is due in part to the sparse measurement sets and short time windows available for most observed variable sources. However, it is also the case that the theoretical models tend to sample possible mechanisms rather than constrain and quantify the range of conditions under which these mechanisms might act. When example outburst lightcurves are produced, they do not seem to match in detail the empirical data in terms of lightcurve profile or duration. Furthermore, the models are not yet fully coupled to the radiative transfer physics necessary to predict observables as a function of wavelength. Observations at infrared wavelengths, both to monitor time-variable emission and to spatially resolve the emitting region, would be extremely desirable to constrain the range of disk radii involved, which in turn has a big impact on theoretical models.

What is needed before the next Protostars and Planets, therefore, is a set of agreed upon measurements that both the theorists and observers can begin to compile. These should include diagnostics of the burst itself such as the rise and decay shapes and timescales. Of particular value will be the full width at half maximum of the burst length as well as the amplitude, or range of potential amplitudes, of the accretion event in both $M_\sun$ yr$^{-1}$ and as a percentage of the steady-state accretion rate. Additionally, a clear understanding of how these values vary with physical parameters is essential for meaningful theoretical comparison. On the observational side, quantifying the physical state of each bursting system, including disk and stellar properties, is essential. Furthermore, given the increasing number of monitoring surveys, statistical approaches will play an ever more important role, allowing for consideration of the time between events of various burst strength and whether there is a quasi-repeatability. Within these surveys, the more routine low-amplitude variable accretion, especially the years-to-decades-long secular component, may be as important to monitor as the rare bursts and outbursts.

\subsection{What Are the Best Practices for Classifying YSO Bursts and Outbursts?}

In the era of all-sky photometric monitoring\index{monitoring} at multiple wavelengths, the literature is full of discoveries of large-amplitude photometric changes. Automated techniques for identifying YSO outbursts in real time, as well as machine-learning efforts to distinguish them from other families of events, are improving. While progress has been made in the latter \citep{richards12,jayasinghe19,vanroestel21, sanchez21}, the aperiodic nature of YSO variability means that these sources lack patterns in their lightcurves that are similar from object to object, and hence YSOs can easily wind up in the ``junk pile.''

Figure \ref{fig:amp_vs_time} highlights that different physical mechanisms may cause variability with similar amplitudes and timescales. Disentangling the physical reason behind the observed photometric changes is not a trivial task. To properly interpret these and assess the significance of each discovery, here we recommend several best practices.

Spectroscopic follow-up is essential\index{spectroscopy}. Spectra are required to understand disk and accretion physics, especially at high resolution, where the gas kinematics can be probed. They allow us to distinguish between FU~Ori outbursts and variability of lesser amplitude. In the optical and near IR, spectra of young stars tell us primarily about gas temperature. For normally accreting pre-main sequence stars, optical/infrared spectra usually show a stellar photosphere in absorption, plus a superposed continuum and emission lines due to the accretion component. As the accretion rate increases, the spectrum is swamped by the accretion component, so it becomes impossible to detect the photospheric absorption component. In extreme outbursts, with very high accretion rates implying very high luminosities, the observed spectrum becomes that of the hot inner disk, which is in absorption.

In addition to spectroscopic follow-up, near- and mid-infrared interferometry will also be powerful approaches to resolving circumstellar disks at all scales, down to nearly the stellar surface. Such techniques may help relate the inner disk structure to the origin of outbursts.

\looseness=1
The importance of spectra illustrates why the variability of optically revealed sources is better understood than that of deeply embedded sources. Photometric and spectroscopic monitoring of the youngest accreting protostars is not yet possible with the cadences available for more evolved ones (or at all), and our understanding of the accretion physics is of necessity less well developed. Spectroscopic capabilities will improve with time, but the requisite information about the most extincted or deeply embedded YSOs may be inaccessible for the foreseeable future.

Besides acquiring and carefully interpreting spectra, additional best practices include investigation into the availability of archival spectroscopic and photometric data. Understanding whether the object has a history of dramatic variability can put the observed burst/outburst in context. It is also important to look at variability across wavelength regimes and whether multiwavelength variability is correlated or uncorrelated. Can a change in extinction be ruled out in favor of a genuine increase in the luminosity?

As noted above, the FU~Ori and EX~Lup categories are not sufficient to cover all cases of bursting YSOs. Detected outbursts, particularly among protostars, have a broad range of amplitudes, durations, and recurrence timescales, many intermediate to these two cases. Thus, when describing individual sources, the best practice may be for authors to use the more general terms ``burst'' or ``outburst'' instead of assigning specific classifications and to carefully describe their selection criteria, when describing searches, or source properties, when describing individual objects. Astronomers are impatient people who are driven to classify. For evaluating the nature of new outbursts, the best practice may be to wait for more data in spite of our impatience. Since the brightening of FU Ori\index[obj]{FU Ori} over 85 years ago, we have made some progress in connecting empirical phenomena to physical processes, but there is still much to learn.

\section{\textbf{SUMMARY}}\label{sec:summary}

We close by emphasizing the main points from our review. First, the critical parameter to determine for each burst is $\int \dot M_{\rm acc} ~dt$: How much mass was accreted? Second, we must look beyond protostellar luminosities to make additional progress on understanding how stars accrete their masses. Third, traditional classification schemes are imprecise, and members assigned to a given category typically display a surprising range of luminosities, lightcurves, and spectra.  Spectroscopy is essential for distinguishing between FU Ori outbursts and variability of lesser amplitude. Fourth, the largest bursts detected in the mid IR generally have smaller amplitudes than the largest bursts detected in the optical. The reason for this and its implications for derived outburst rates are topics for ongoing study. Finally, to further develop our understanding of stellar mass assembly via variable accretion, observers need to carefully monitor individual (out)bursts while also undertaking a systematic investigation into the full range of accretion variability phenomena, while theorists must continue to work toward more precise observational diagnostics.

\acknowledgements
This project has received funding from the European Research Council (ERC) under the European Union's Horizon 2020 research and innovation programme under grant agreement No.\ 716155 (SACCRED). DJ is supported by NRC Canada and by an NSERC Discovery Grant. GJH is supported by the National Natural Science Foundation of China grant 12173003.

\bigskip

{\small
\baselineskip=11pt
\bibliographystyle{pp7}
\bibliography{Chapter10.bib}

\begin{thebibliography}{235}
\parskip=0pt \itemsep=0pt \small \baselineskip=11pt
\providecommand{\natexlab}[1]{#1}

\bibitem[\protect\astroncite{\emph{{{\'A}brah{\'a}m} et~al.}}{2009}]{abraham09}
{{\'A}brah{\'a}m} P. et~al., 2009 \emph{\nat}, \emph{459}, 7244, 224.

\bibitem[\protect\astroncite{\emph{{{\'A}brah{\'a}m} et~al.}}{2019}]{abraham19}
{{\'A}brah{\'a}m} P. et~al., 2019 \emph{\apj}, \emph{887}, 2, 156.

\bibitem[\protect\astroncite{\emph{{Acosta-Pulido}
  et~al.}}{2007}]{acosta-pulido07}
{Acosta-Pulido} J.~A. et~al., 2007 \emph{\aj}, \emph{133}, 5, 2020.

\bibitem[\protect\astroncite{\emph{{Alencar} et~al.}}{2012}]{alencar12}
{Alencar} S.~H.~P. et~al., 2012 \emph{\aap}, \emph{541}, A116.

\bibitem[\protect\astroncite{\emph{{Anderl} et~al.}}{2020}]{anderl20}
{Anderl} S. et~al., 2020 \emph{\aap}, \emph{643}, A123.

\bibitem[\protect\astroncite{\emph{{Ansdell} et~al.}}{2016}]{ansdell16}
{Ansdell} M. et~al., 2016 \emph{\apj}, \emph{816}, 2, 69.

\bibitem[\protect\astroncite{\emph{{Armitage}}}{2015}]{armitage15}
{Armitage} P.~J., 2015 \emph{arXiv e-prints}, arXiv:1509.06382.

\bibitem[\protect\astroncite{\emph{{Armitage}}}{2016}]{armitage16}
{Armitage} P.~J., 2016 \emph{\apjl}, \emph{833}, 2, L15.

\bibitem[\protect\astroncite{\emph{{Armitage} et~al.}}{2001}]{armitage01}
{Armitage} P.~J. et~al., 2001 \emph{\mnras}, \emph{324}, 3, 705.

\bibitem[\protect\astroncite{\emph{{Aspin} and {Reipurth}}}{2003}]{aspin03}
{Aspin} C. and {Reipurth} B., 2003 \emph{\aj}, \emph{126}, 6, 2936.

\bibitem[\protect\astroncite{\emph{{Aspin} et~al.}}{2010}]{aspin10}
{Aspin} C. et~al., 2010 \emph{\apjl}, \emph{719}, 1, L50.

\bibitem[\protect\astroncite{\emph{{Audard} et~al.}}{2014}]{audard14}
{Audard} M. et~al., 2014 \emph{Protostars and Planets VI} (H.~{Beuther}, R.~S.
  {Klessen}, C.~P. {Dullemond}, and T.~{Henning}), p. 387.

\bibitem[\protect\astroncite{\emph{{Bae} et~al.}}{2014}]{bae14}
{Bae} J. et~al., 2014 \emph{\apj}, \emph{795}, 1, 61.

\bibitem[\protect\astroncite{\emph{{Bae} et~al.}}{2015}]{bae15}
{Bae} J. et~al., 2015 \emph{\apj}, \emph{805}, 1, 15.

\bibitem[\protect\astroncite{\emph{{Baek} et~al.}}{2020}]{baek20}
{Baek} G. et~al., 2020 \emph{\apj}, \emph{895}, 1, 27.

\bibitem[\protect\astroncite{\emph{{Bai}}}{2011}]{bai11}
{Bai} X.-N., 2011 \emph{\apj}, \emph{739}, 1, 50.

\bibitem[\protect\astroncite{\emph{{Bai}}}{2017}]{bai17}
{Bai} X.-N., 2017 \emph{\apj}, \emph{845}, 1, 75.

\bibitem[\protect\astroncite{\emph{{Banzatti} et~al.}}{2015}]{banzatti15}
{Banzatti} A. et~al., 2015 \emph{\apjl}, \emph{798}, 1, L16.

\bibitem[\protect\astroncite{\emph{{Benisty} et~al.}}{2010}]{benisty10}
{Benisty} M. et~al., 2010 \emph{\aap}, \emph{517}, L3.

\bibitem[\protect\astroncite{\emph{{Bianchi} et~al.}}{2020}]{bianchi20}
{Bianchi} E. et~al., 2020 \emph{\mnras}, \emph{498}, 1, L87.

\bibitem[\protect\astroncite{\emph{{Blinova} et~al.}}{2016}]{blinova16}
{Blinova} A.~A. et~al., 2016 \emph{\mnras}, \emph{459}, 3, 2354.

\bibitem[\protect\astroncite{\emph{{Bouvier} et~al.}}{2007}]{bouvier07}
{Bouvier} J. et~al., 2007 \emph{\aap}, \emph{463}, 3, 1017.

\bibitem[\protect\astroncite{\emph{{Bouvier} et~al.}}{2013}]{bouvier13}
{Bouvier} J. et~al., 2013 \emph{\aap}, \emph{557}, A77.

\bibitem[\protect\astroncite{\emph{{Brice{\~n}o} et~al.}}{2004}]{briceno04}
{Brice{\~n}o} C. et~al., 2004 \emph{\apjl}, \emph{606}, 2, L123.

\bibitem[\protect\astroncite{\emph{{Burns} et~al.}}{2016}]{burns16}
{Burns} R.~A. et~al., 2016 \emph{\mnras}, \emph{460}, 1, 283.

\bibitem[\protect\astroncite{\emph{{Burns} et~al.}}{2020}]{burns20}
{Burns} R.~A. et~al., 2020 \emph{Nature Astronomy}, \emph{4}, 506.

\bibitem[\protect\astroncite{\emph{{Campbell-White}
  et~al.}}{2021}]{campbellwhite21}
{Campbell-White} J. et~al., 2021 \emph{\mnras}, \emph{507}, 3, 3331.

\bibitem[\protect\astroncite{\emph{{Caratti o Garatti}
  et~al.}}{2011}]{carattiogaratti11}
{Caratti o Garatti} A. et~al., 2011 \emph{\aap}, \emph{526}, L1.

\bibitem[\protect\astroncite{\emph{{Caratti o Garatti}
  et~al.}}{2017}]{caratti-o-garatti17}
{Caratti o Garatti} A. et~al., 2017 \emph{Nature Physics}, \emph{13}, 3, 276.

\bibitem[\protect\astroncite{\emph{{Chen} et~al.}}{2021}]{chen21}
{Chen} Z. et~al., 2021 \emph{\apj}, \emph{922}, 1, 90.

\bibitem[\protect\astroncite{\emph{{Cieza} et~al.}}{2018}]{cieza18}
{Cieza} L.~A. et~al., 2018 \emph{\mnras}, \emph{474}, 4, 4347.

\bibitem[\protect\astroncite{\emph{{Clarke} et~al.}}{2005}]{clarke05}
{Clarke} C. et~al., 2005 \emph{\mnras}, \emph{361}, 3, 942.

\bibitem[\protect\astroncite{\emph{{Cody} and {Hillenbrand}}}{2010}]{cody10}
{Cody} A.~M. and {Hillenbrand} L.~A., 2010 \emph{\apjs}, \emph{191}, 2, 389.

\bibitem[\protect\astroncite{\emph{{Cody} et~al.}}{2014}]{cody14}
{Cody} A.~M. et~al., 2014 \emph{\aj}, \emph{147}, 4, 82.

\bibitem[\protect\astroncite{\emph{{Cody} et~al.}}{2017}]{cody17}
{Cody} A.~M. et~al., 2017 \emph{\apj}, \emph{836}, 1, 41.

\bibitem[\protect\astroncite{\emph{{Connelley} and
  {Reipurth}}}{2018}]{connelley18}
{Connelley} M.~S. and {Reipurth} B., 2018 \emph{\apj}, \emph{861}, 2, 145.

\bibitem[\protect\astroncite{\emph{{Contreras Pe{\~n}a}
  et~al.}}{2017{\natexlab{a}}}]{contreraspena17a}
{Contreras Pe{\~n}a} C. et~al., 2017{\natexlab{a}} \emph{\mnras}, \emph{465},
  3, 3011.

\bibitem[\protect\astroncite{\emph{{Contreras Pe{\~n}a}
  et~al.}}{2017{\natexlab{b}}}]{contreraspena17b}
{Contreras Pe{\~n}a} C. et~al., 2017{\natexlab{b}} \emph{\mnras}, \emph{465},
  3, 3039.

\bibitem[\protect\astroncite{\emph{{Contreras Pe{\~n}a}
  et~al.}}{2019}]{contreraspena19}
{Contreras Pe{\~n}a} C. et~al., 2019 \emph{\mnras}, \emph{486}, 4, 4590.

\bibitem[\protect\astroncite{\emph{{Contreras Pe{\~n}a}
  et~al.}}{2020}]{contreraspena20}
{Contreras Pe{\~n}a} C. et~al., 2020 \emph{\mnras}, \emph{495}, 4, 3614.

\bibitem[\protect\astroncite{\emph{{Costigan} et~al.}}{2014}]{costigan14}
{Costigan} G. et~al., 2014 \emph{\mnras}, \emph{440}, 4, 3444.

\bibitem[\protect\astroncite{\emph{{Covey} et~al.}}{2011}]{covey11}
{Covey} K.~R. et~al., 2011 \emph{\aj}, \emph{141}, 2, 40.

\bibitem[\protect\astroncite{\emph{{Covey} et~al.}}{2021}]{covey21}
{Covey} K.~R. et~al., 2021 \emph{\aj}, \emph{161}, 2, 61.

\bibitem[\protect\astroncite{\emph{{Dahm} and {Hillenbrand}}}{2020}]{dahm20}
{Dahm} S.~E. and {Hillenbrand} L.~A., 2020 \emph{\aj}, \emph{160}, 6, 278.

\bibitem[\protect\astroncite{\emph{{D'Angelo} and {Spruit}}}{2012}]{dangelo12}
{D'Angelo} C.~R. and {Spruit} H.~C., 2012 \emph{\mnras}, \emph{420}, 1, 416.

\bibitem[\protect\astroncite{\emph{{Davies}}}{2019}]{davies19}
{Davies} C.~L., 2019 \emph{\mnras}, \emph{484}, 2, 1926.

\bibitem[\protect\astroncite{\emph{{Desch} and {Turner}}}{2015}]{desch15}
{Desch} S.~J. and {Turner} N.~J., 2015 \emph{\apj}, \emph{811}, 2, 156.

\bibitem[\protect\astroncite{\emph{{Dodin} et~al.}}{2016}]{dodin16}
{Dodin} A.~V. et~al., 2016 \emph{Astronomy Letters}, \emph{42}, 1, 29.

\bibitem[\protect\astroncite{\emph{{Dunham} and {Vorobyov}}}{2012}]{dunham12}
{Dunham} M.~M. and {Vorobyov} E.~I., 2012 \emph{\apj}, \emph{747}, 1, 52.

\bibitem[\protect\astroncite{\emph{{Dunham} et~al.}}{2014}]{dunham14}
{Dunham} M.~M. et~al., 2014 \emph{Protostars and Planets VI} (H.~{Beuther},
  R.~S. {Klessen}, C.~P. {Dullemond}, and T.~{Henning}), p. 195.

\bibitem[\protect\astroncite{\emph{{Dunham} et~al.}}{2015}]{dunham15}
{Dunham} M.~M. et~al., 2015 \emph{\apjs}, \emph{220}, 1, 11.

\bibitem[\protect\astroncite{\emph{{Elias}}}{1978}]{elias78}
{Elias} J.~H., 1978 \emph{\apj}, \emph{223}, 859.

\bibitem[\protect\astroncite{\emph{{Ellerbroek} et~al.}}{2014}]{ellerbroek14}
{Ellerbroek} L.~E. et~al., 2014 \emph{\aap}, \emph{563}, A87.

\bibitem[\protect\astroncite{\emph{{Enoch} et~al.}}{2009}]{enoch09}
{Enoch} M.~L. et~al., 2009 \emph{\apj}, \emph{692}, 2, 973.

\bibitem[\protect\astroncite{\emph{{Espaillat} et~al.}}{2011}]{espaillat11}
{Espaillat} C. et~al., 2011 \emph{\apj}, \emph{728}, 1, 49.

\bibitem[\protect\astroncite{\emph{{Espaillat} et~al.}}{2021}]{espaillat21}
{Espaillat} C.~C. et~al., 2021 \emph{\nat}, \emph{597}, 7874, 41.

\bibitem[\protect\astroncite{\emph{{Evans} et~al.}}{2009}]{evans09}
{Evans} Neal~J. I. et~al., 2009 \emph{\apjs}, \emph{181}, 2, 321.

\bibitem[\protect\astroncite{\emph{{Fischer} and
  {Hillenbrand}}}{2017}]{fischer17b}
{Fischer} W.~J. and {Hillenbrand} L., 2017 \emph{The Astronomer's Telegram},
  \emph{9969}, 1.

\bibitem[\protect\astroncite{\emph{{Fischer} et~al.}}{2012}]{fischer12}
{Fischer} W.~J. et~al., 2012 \emph{\apj}, \emph{756}, 1, 99.

\bibitem[\protect\astroncite{\emph{{Fischer} et~al.}}{2017}]{fischer17}
{Fischer} W.~J. et~al., 2017 \emph{\apj}, \emph{840}, 2, 69.

\bibitem[\protect\astroncite{\emph{{Fischer} et~al.}}{2019}]{fischer19}
{Fischer} W.~J. et~al., 2019 \emph{\apj}, \emph{872}, 2, 183.

\bibitem[\protect\astroncite{\emph{{Flaherty} et~al.}}{2016}]{flaherty16}
{Flaherty} K.~M. et~al., 2016 \emph{\apj}, \emph{833}, 1, 104.

\bibitem[\protect\astroncite{\emph{{Flock} et~al.}}{2017}]{flock17}
{Flock} M. et~al., 2017 \emph{\apj}, \emph{835}, 2, 230.

\bibitem[\protect\astroncite{\emph{{Frank} et~al.}}{2014}]{frank14}
{Frank} A. et~al., 2014 \emph{Protostars and Planets VI} (H.~{Beuther}, R.~S.
  {Klessen}, C.~P. {Dullemond}, and T.~{Henning}), p. 451.

\bibitem[\protect\astroncite{\emph{{Frimann} et~al.}}{2017}]{frimann17}
{Frimann} S. et~al., 2017 \emph{\aap}, \emph{602}, A120.

\bibitem[\protect\astroncite{\emph{{Gammie}}}{1996}]{gammie96}
{Gammie} C.~F., 1996 \emph{\apj}, \emph{457}, 355.

\bibitem[\protect\astroncite{\emph{{Garufi} et~al.}}{2019}]{garufi19}
{Garufi} A. et~al., 2019 \emph{\aap}, \emph{628}, A68.

\bibitem[\protect\astroncite{\emph{{Giannini} et~al.}}{2020}]{giannini20}
{Giannini} T. et~al., 2020 \emph{\aap}, \emph{637}, A83.

\bibitem[\protect\astroncite{\emph{{Greene} et~al.}}{2008}]{greene08}
{Greene} T.~P. et~al., 2008 \emph{\aj}, \emph{135}, 4, 1421.

\bibitem[\protect\astroncite{\emph{{Gregory} and {Donati}}}{2011}]{gregory11}
{Gregory} S.~G. and {Donati} J.~F., 2011 \emph{Astronomische Nachrichten},
  \emph{332}, 1027.

\bibitem[\protect\astroncite{\emph{{Grosso} et~al.}}{2020}]{grosso20}
{Grosso} N. et~al., 2020 \emph{\aap}, \emph{638}, L4.

\bibitem[\protect\astroncite{\emph{{Guarcello} et~al.}}{2019}]{guarcello19}
{Guarcello} M.~G. et~al., 2019 \emph{\aap}, \emph{628}, A74.

\bibitem[\protect\astroncite{\emph{{Guo} et~al.}}{2018}]{guo18}
{Guo} Z. et~al., 2018 \emph{\apj}, \emph{852}, 1, 56.

\bibitem[\protect\astroncite{\emph{{Guo} et~al.}}{2020}]{guo20}
{Guo} Z. et~al., 2020 \emph{\mnras}, \emph{492}, 1, 294.

\bibitem[\protect\astroncite{\emph{{Hackstein} et~al.}}{2015}]{hackstein15}
{Hackstein} M. et~al., 2015 \emph{\aap}, \emph{582}, L12.

\bibitem[\protect\astroncite{\emph{{Harris} and {Beichman}}}{1980}]{harris80}
{Harris} S. and {Beichman} C., 1980 {The formation of a T Tauri star:
  Observations of the infrared source in L 1551}, NASA STI/Recon Technical
  Report N.

\bibitem[\protect\astroncite{\emph{{Hartmann} and {Kenyon}}}{1985}]{hartmann85}
{Hartmann} L. and {Kenyon} S.~J., 1985 \emph{\apj}, \emph{299}, 462.

\bibitem[\protect\astroncite{\emph{{Hartmann} and {Kenyon}}}{1996}]{hartmann96}
{Hartmann} L. and {Kenyon} S.~J., 1996 \emph{\araa}, \emph{34}, 207.

\bibitem[\protect\astroncite{\emph{{Hartmann} et~al.}}{2016}]{hartmann16}
{Hartmann} L. et~al., 2016 \emph{\araa}, \emph{54}, 135.

\bibitem[\protect\astroncite{\emph{{Heiderman} and
  {Evans}}}{2015}]{heiderman15}
{Heiderman} A. and {Evans} Neal~J. I., 2015 \emph{\apj}, \emph{806}, 2, 231.

\bibitem[\protect\astroncite{\emph{{Herbig}}}{1989}]{herbig89}
{Herbig} G.~H., 1989 \emph{European Southern Observatory Conference and
  Workshop Proceedings}, vol.~33 of \emph{European Southern Observatory
  Conference and Workshop Proceedings}, pp. 233--246.

\bibitem[\protect\astroncite{\emph{{Herbig}}}{1990}]{herbig90}
{Herbig} G.~H., 1990 \emph{\apj}, \emph{360}, 639.

\bibitem[\protect\astroncite{\emph{{Herbig}}}{2008}]{herbig08}
{Herbig} G.~H., 2008 \emph{\aj}, \emph{135}, 2, 637.

\bibitem[\protect\astroncite{\emph{{Herbig} et~al.}}{2003}]{herbig03}
{Herbig} G.~H. et~al., 2003 \emph{\apj}, \emph{595}, 1, 384.

\bibitem[\protect\astroncite{\emph{{Herbst} et~al.}}{1994}]{herbst94}
{Herbst} W. et~al., 1994 \emph{\aj}, \emph{108}, 1906.

\bibitem[\protect\astroncite{\emph{{Herczeg} et~al.}}{2016}]{herczeg16}
{Herczeg} G.~J. et~al., 2016 \emph{\apj}, \emph{831}, 2, 133.

\bibitem[\protect\astroncite{\emph{{Herczeg} et~al.}}{2019}]{herczeg19}
{Herczeg} G.~J. et~al., 2019 \emph{\apj}, \emph{878}, 2, 111.

\bibitem[\protect\astroncite{\emph{{Hillenbrand} and
  {Findeisen}}}{2015}]{hillenbrand15}
{Hillenbrand} L.~A. and {Findeisen} K.~P., 2015 \emph{\apj}, \emph{808}, 1, 68.

\bibitem[\protect\astroncite{\emph{{Hillenbrand} and
  {Rodriguez}}}{2022}]{hillenbrand22}
{Hillenbrand} L.~A. and {Rodriguez} A.~C., 2022 \emph{Research Notes of the
  American Astronomical Society}, \emph{6}, 1, 6.

\bibitem[\protect\astroncite{\emph{{Hillenbrand} et~al.}}{2013}]{hillenbrand13}
{Hillenbrand} L.~A. et~al., 2013 \emph{\aj}, \emph{145}, 3, 59.

\bibitem[\protect\astroncite{\emph{{Hillenbrand} et~al.}}{2018}]{hillenbrand18}
{Hillenbrand} L.~A. et~al., 2018 \emph{\apj}, \emph{869}, 2, 146.

\bibitem[\protect\astroncite{\emph{{Hillenbrand}
  et~al.}}{2019{\natexlab{a}}}]{hillenbrand19b}
{Hillenbrand} L.~A. et~al., 2019{\natexlab{a}} \emph{\aj}, \emph{158}, 6, 240.

\bibitem[\protect\astroncite{\emph{{Hillenbrand}
  et~al.}}{2019{\natexlab{b}}}]{hillenbrand19a}
{Hillenbrand} L.~A. et~al., 2019{\natexlab{b}} \emph{\apj}, \emph{874}, 1, 82.

\bibitem[\protect\astroncite{\emph{{Hodapp} et~al.}}{2012}]{hodapp12}
{Hodapp} K.~W. et~al., 2012 \emph{\apj}, \emph{744}, 56.

\bibitem[\protect\astroncite{\emph{{Hodapp} et~al.}}{2019}]{hodapp19}
{Hodapp} K.~W. et~al., 2019 \emph{\aj}, \emph{158}, 6, 241.

\bibitem[\protect\astroncite{\emph{{Hodapp} et~al.}}{2020}]{hodapp20}
{Hodapp} K.~W. et~al., 2020 \emph{\aj}, \emph{160}, 4, 164.

\bibitem[\protect\astroncite{\emph{{Hsieh} et~al.}}{2019}]{hsieh19}
{Hsieh} T.-H. et~al., 2019 \emph{\apj}, \emph{884}, 2, 149.

\bibitem[\protect\astroncite{\emph{{Hunter} et~al.}}{2017}]{hunter17}
{Hunter} T.~R. et~al., 2017 \emph{\apjl}, \emph{837}, 2, L29.

\bibitem[\protect\astroncite{\emph{{Jayasinghe} et~al.}}{2019}]{jayasinghe19}
{Jayasinghe} T. et~al., 2019 \emph{\mnras}, \emph{486}, 2, 1907.

\bibitem[\protect\astroncite{\emph{{Johnstone} et~al.}}{2013}]{johnstone13}
{Johnstone} D. et~al., 2013 \emph{\apj}, \emph{765}, 133.

\bibitem[\protect\astroncite{\emph{{Johnstone} et~al.}}{2018}]{johnstone18}
{Johnstone} D. et~al., 2018 \emph{\apj}, \emph{854}, 1, 31.

\bibitem[\protect\astroncite{\emph{{J{\o}rgensen} et~al.}}{2015}]{jorgensen15}
{J{\o}rgensen} J.~K. et~al., 2015 \emph{\aap}, \emph{579}, A23.

\bibitem[\protect\astroncite{\emph{{J{\o}rgensen} et~al.}}{2020}]{jorgensen20}
{J{\o}rgensen} J.~K. et~al., 2020 \emph{\araa}, \emph{58}, 727.

\bibitem[\protect\astroncite{\emph{{Juh{\'a}sz} et~al.}}{2012}]{juhasz12}
{Juh{\'a}sz} A. et~al., 2012 \emph{\apj}, \emph{744}, 2, 118.

\bibitem[\protect\astroncite{\emph{{Kadam} et~al.}}{2019}]{kadam19}
{Kadam} K. et~al., 2019 \emph{\apj}, \emph{882}, 2, 96.

\bibitem[\protect\astroncite{\emph{{Kadam} et~al.}}{2020}]{kadam20}
{Kadam} K. et~al., 2020 \emph{\apj}, \emph{895}, 1, 41.

\bibitem[\protect\astroncite{\emph{{Kenyon} and {Hartmann}}}{1995}]{kenyon95}
{Kenyon} S.~J. and {Hartmann} L., 1995 \emph{\apjs}, \emph{101}, 117.

\bibitem[\protect\astroncite{\emph{{Kenyon} et~al.}}{1988}]{kenyon88}
{Kenyon} S.~J. et~al., 1988 \emph{\apj}, \emph{325}, 231.

\bibitem[\protect\astroncite{\emph{{Kenyon} et~al.}}{1990}]{kenyon90}
{Kenyon} S.~J. et~al., 1990 \emph{\aj}, \emph{99}, 869.

\bibitem[\protect\astroncite{\emph{{Kenyon} et~al.}}{1994}]{kenyon94}
{Kenyon} S.~J. et~al., 1994 \emph{\aj}, \emph{108}, 251.

\bibitem[\protect\astroncite{\emph{{Kenyon} et~al.}}{2000}]{kenyon00}
{Kenyon} S.~J. et~al., 2000 \emph{\apj}, \emph{531}, 2, 1028.

\bibitem[\protect\astroncite{\emph{{Kesseli} et~al.}}{2016}]{kesseli16}
{Kesseli} A.~Y. et~al., 2016 \emph{\apj}, \emph{828}, 1, 42.

\bibitem[\protect\astroncite{\emph{{Kopatskaya} et~al.}}{2013}]{kopatskaya13}
{Kopatskaya} E.~N. et~al., 2013 \emph{\mnras}, \emph{434}, 1, 38.

\bibitem[\protect\astroncite{\emph{{K{\'o}sp{\'a}l} et~al.}}{2007}]{kospal07}
{K{\'o}sp{\'a}l} {\'A}. et~al., 2007 \emph{\aap}, \emph{470}, 1, 211.

\bibitem[\protect\astroncite{\emph{{K{\'o}sp{\'a}l} et~al.}}{2008}]{kospal08}
{K{\'o}sp{\'a}l} {\'A}. et~al., 2008 \emph{\mnras}, \emph{383}, 3, 1015.

\bibitem[\protect\astroncite{\emph{{K{\'o}sp{\'a}l} et~al.}}{2011}]{kospal11}
{K{\'o}sp{\'a}l} {\'A}. et~al., 2011 \emph{\aap}, \emph{527}, A133.

\bibitem[\protect\astroncite{\emph{{K{\'o}sp{\'a}l} et~al.}}{2013}]{kospal13}
{K{\'o}sp{\'a}l} {\'A}. et~al., 2013 \emph{\aap}, \emph{551}, A62.

\bibitem[\protect\astroncite{\emph{{K{\'o}sp{\'a}l} et~al.}}{2016}]{kospal16}
{K{\'o}sp{\'a}l} {\'A}. et~al., 2016 \emph{\aap}, \emph{596}, A52.

\bibitem[\protect\astroncite{\emph{{K{\'o}sp{\'a}l}
  et~al.}}{2017{\natexlab{a}}}]{kospal17}
{K{\'o}sp{\'a}l} {\'A}. et~al., 2017{\natexlab{a}} \emph{\aap}, \emph{597},
  L10.

\bibitem[\protect\astroncite{\emph{{K{\'o}sp{\'a}l}
  et~al.}}{2017{\natexlab{b}}}]{kospal17b}
{K{\'o}sp{\'a}l} {\'A}. et~al., 2017{\natexlab{b}} \emph{\apj}, \emph{843}, 1,
  45.

\bibitem[\protect\astroncite{\emph{{K{\'o}sp{\'a}l} et~al.}}{2021}]{kospal21}
{K{\'o}sp{\'a}l} {\'A}. et~al., 2021 \emph{\apjs}, \emph{256}, 2, 30.

\bibitem[\protect\astroncite{\emph{{Kraus} et~al.}}{2016}]{kraus16}
{Kraus} S. et~al., 2016 \emph{\mnras}, \emph{462}, 1, L61.

\bibitem[\protect\astroncite{\emph{{Kristensen} and
  {Dunham}}}{2018}]{kristensen18}
{Kristensen} L.~E. and {Dunham} M.~M., 2018 \emph{\aap}, \emph{618}, A158.

\bibitem[\protect\astroncite{\emph{{Kryukova} et~al.}}{2012}]{kryukova12}
{Kryukova} E. et~al., 2012 \emph{\aj}, \emph{144}, 2, 31.

\bibitem[\protect\astroncite{\emph{{Kuffmeier} et~al.}}{2018}]{kuffmeier18}
{Kuffmeier} M. et~al., 2018 \emph{\mnras}, \emph{475}, 2, 2642.

\bibitem[\protect\astroncite{\emph{{Labdon} et~al.}}{2021}]{labdon21}
{Labdon} A. et~al., 2021 \emph{\aap}, \emph{646}, A102.

\bibitem[\protect\astroncite{\emph{{Lee}}}{2007}]{lee07}
{Lee} J.-E., 2007 \emph{Journal of Korean Astronomical Society}, \emph{40}, 83.

\bibitem[\protect\astroncite{\emph{{Lee} et~al.}}{2019}]{lee19}
{Lee} J.-E. et~al., 2019 \emph{Nature Astronomy}, \emph{3}, 314.

\bibitem[\protect\astroncite{\emph{{Lee} et~al.}}{2021{\natexlab{a}}}]{leejl21}
{Lee} J.-E. et~al., 2021{\natexlab{a}} \emph{\apjl}, \emph{916}, 2, L20.

\bibitem[\protect\astroncite{\emph{{Lee} et~al.}}{2020}]{lee20}
{Lee} Y.-H. et~al., 2020 \emph{\apj}, \emph{903}, 1, 5.

\bibitem[\protect\astroncite{\emph{{Lee} et~al.}}{2021{\natexlab{b}}}]{leeyh21}
{Lee} Y.-H. et~al., 2021{\natexlab{b}} \emph{\apj}, \emph{920}, 2, 119.

\bibitem[\protect\astroncite{\emph{{Lesur} et~al.}}{2015}]{lesur15}
{Lesur} G. et~al., 2015 \emph{\aap}, \emph{582}, L9.

\bibitem[\protect\astroncite{\emph{{Lesur}}}{2021}]{lesur21}
{Lesur} G. R.~J., 2021 \emph{\aap}, \emph{650}, A35.

\bibitem[\protect\astroncite{\emph{{Liu} et~al.}}{2018}]{liu18}
{Liu} H.~B. et~al., 2018 \emph{\aap}, \emph{612}, A54.

\bibitem[\protect\astroncite{\emph{{Loomis} et~al.}}{2017}]{loomis17}
{Loomis} R.~A. et~al., 2017 \emph{\apj}, \emph{840}, 1, 23.

\bibitem[\protect\astroncite{\emph{{Lorenzetti} et~al.}}{2009}]{lorenzetti09}
{Lorenzetti} D. et~al., 2009 \emph{\apj}, \emph{693}, 2, 1056.

\bibitem[\protect\astroncite{\emph{{Lucas} et~al.}}{2020}]{lucas20}
{Lucas} P.~W. et~al., 2020 \emph{\mnras}, \emph{499}, 2, 1805.

\bibitem[\protect\astroncite{\emph{{MacFarlane} et~al.}}{2019}]{macfarlane19a}
{MacFarlane} B. et~al., 2019 \emph{\mnras}, \emph{487}, 4, 5106.

\bibitem[\protect\astroncite{\emph{{MacLeod} et~al.}}{2018}]{macleod18}
{MacLeod} G.~C. et~al., 2018 \emph{\mnras}, \emph{478}, 1, 1077.

\bibitem[\protect\astroncite{\emph{{Magakian} et~al.}}{2019}]{magakian19}
{Magakian} T.~Y. et~al., 2019 \emph{\aap}, \emph{625}, A13.

\bibitem[\protect\astroncite{\emph{{Mairs} et~al.}}{2017}]{mairs17b}
{Mairs} S. et~al., 2017 \emph{\apj}, \emph{849}, 2, 107.

\bibitem[\protect\astroncite{\emph{{Maksimova} et~al.}}{2020}]{maksimova20}
{Maksimova} L.~A. et~al., 2020 \emph{Astronomy Reports}, \emph{64}, 10, 815.

\bibitem[\protect\astroncite{\emph{{Manara} et~al.}}{2019}]{manara19}
{Manara} C.~F. et~al., 2019 \emph{\aap}, \emph{631}, L2.

\bibitem[\protect\astroncite{\emph{{Manara} et~al.}}{2021}]{manara21}
{Manara} C.~F. et~al., 2021 \emph{\aap}, \emph{650}, A196.

\bibitem[\protect\astroncite{\emph{{Martin} and {Lubow}}}{2011}]{martin11}
{Martin} R.~G. and {Lubow} S.~H., 2011 \emph{\apjl}, \emph{740}, 1, L6.

\bibitem[\protect\astroncite{\emph{{McGinnis} et~al.}}{2015}]{mcginnis15}
{McGinnis} P.~T. et~al., 2015 \emph{\aap}, \emph{577}, A11.

\bibitem[\protect\astroncite{\emph{{McKee} and {Offner}}}{2010}]{mckee10}
{McKee} C.~F. and {Offner} S. S.~R., 2010 \emph{\apj}, \emph{716}, 1, 167.

\bibitem[\protect\astroncite{\emph{{Mendigut{\'\i}a}
  et~al.}}{2013}]{mendigutia13}
{Mendigut{\'\i}a} I. et~al., 2013 \emph{\apj}, \emph{776}, 1, 44.

\bibitem[\protect\astroncite{\emph{{Miller} et~al.}}{2011}]{miller11}
{Miller} A.~A. et~al., 2011 \emph{\apj}, \emph{730}, 2, 80.

\bibitem[\protect\astroncite{\emph{{Molyarova} et~al.}}{2018}]{molyarova18}
{Molyarova} T. et~al., 2018 \emph{\apj}, \emph{866}, 1, 46.

\bibitem[\protect\astroncite{\emph{{Morales-Calder{\'o}n}
  et~al.}}{2011}]{morales-calderon11}
{Morales-Calder{\'o}n} M. et~al., 2011 \emph{\apj}, \emph{733}, 1, 50.

\bibitem[\protect\astroncite{\emph{{Mosoni} et~al.}}{2013}]{mosoni13}
{Mosoni} L. et~al., 2013 \emph{\aap}, \emph{552}, A62.

\bibitem[\protect\astroncite{\emph{{Mu{\~n}oz} and {Lai}}}{2016}]{munoz16}
{Mu{\~n}oz} D.~J. and {Lai} D., 2016 \emph{\apj}, \emph{827}, 1, 43.

\bibitem[\protect\astroncite{\emph{{Mulders} et~al.}}{2017}]{mulders17}
{Mulders} G.~D. et~al., 2017 \emph{\apj}, \emph{847}, 1, 31.

\bibitem[\protect\astroncite{\emph{{Muzerolle} et~al.}}{2013}]{muzerolle13}
{Muzerolle} J. et~al., 2013 \emph{\nat}, \emph{493}, 7432, 378.

\bibitem[\protect\astroncite{\emph{{Myers}}}{2009}]{myers09}
{Myers} P.~C., 2009 \emph{\apj}, \emph{706}, 2, 1341.

\bibitem[\protect\astroncite{\emph{{Myers}}}{2010}]{myers10}
{Myers} P.~C., 2010 \emph{\apj}, \emph{714}, 2, 1280.

\bibitem[\protect\astroncite{\emph{{Ninan} et~al.}}{2013}]{ninan13}
{Ninan} J.~P. et~al., 2013 \emph{\apj}, \emph{778}, 2, 116.

\bibitem[\protect\astroncite{\emph{{Ninan} et~al.}}{2015}]{ninan15}
{Ninan} J.~P. et~al., 2015 \emph{\apj}, \emph{815}, 1, 4.

\bibitem[\protect\astroncite{\emph{{Nony} et~al.}}{2020}]{nony20}
{Nony} T. et~al., 2020 \emph{\aap}, \emph{636}, A38.

\bibitem[\protect\astroncite{\emph{{Offner} and {McKee}}}{2011}]{offner11}
{Offner} S. S.~R. and {McKee} C.~F., 2011 \emph{\apj}, \emph{736}, 1, 53.

\bibitem[\protect\astroncite{\emph{{Okoda} et~al.}}{2018}]{okoda18}
{Okoda} Y. et~al., 2018 \emph{\apjl}, \emph{864}, 2, L25.

\bibitem[\protect\astroncite{\emph{{Park} et~al.}}{2020}]{park20}
{Park} S. et~al., 2020 \emph{\apj}, \emph{900}, 1, 36.

\bibitem[\protect\astroncite{\emph{{Park}
  et~al.}}{2021{\natexlab{a}}}]{park21b}
{Park} S. et~al., 2021{\natexlab{a}} \emph{\apj}, \emph{923}, 2, 171.

\bibitem[\protect\astroncite{\emph{{Park} et~al.}}{2021{\natexlab{b}}}]{park21}
{Park} W. et~al., 2021{\natexlab{b}} \emph{\apj}, \emph{920}, 2, 132.

\bibitem[\protect\astroncite{\emph{{Pavlyuchenkov}
  et~al.}}{2020}]{pavlyuchenkov20}
{Pavlyuchenkov} Y.~N. et~al., 2020 \emph{Astronomy Reports}, \emph{64}, 1, 1.

\bibitem[\protect\astroncite{\emph{{Petrov} et~al.}}{2015}]{petrov15}
{Petrov} P.~P. et~al., 2015 \emph{\aap}, \emph{577}, A73.

\bibitem[\protect\astroncite{\emph{{Plunkett} et~al.}}{2015}]{plunkett15}
{Plunkett} A.~L. et~al., 2015 \emph{\nat}, \emph{527}, 7576, 70.

\bibitem[\protect\astroncite{\emph{{Pouilly} et~al.}}{2021}]{pouilly21}
{Pouilly} K. et~al., 2021 \emph{\aap}, \emph{656}, A50.

\bibitem[\protect\astroncite{\emph{{Rab} et~al.}}{2017}]{rab17}
{Rab} C. et~al., 2017 \emph{\aap}, \emph{604}, A15.

\bibitem[\protect\astroncite{\emph{{Rebull} et~al.}}{2014}]{rebull14}
{Rebull} L.~M. et~al., 2014 \emph{\aj}, \emph{148}, 5, 92.

\bibitem[\protect\astroncite{\emph{{Rebull} et~al.}}{2015}]{rebull15}
{Rebull} L.~M. et~al., 2015 \emph{\aj}, \emph{150}, 6, 175.

\bibitem[\protect\astroncite{\emph{{Reipurth}}}{1989}]{reipurth89}
{Reipurth} B., 1989 \emph{\nat}, \emph{340}, 6228, 42.

\bibitem[\protect\astroncite{\emph{{Reipurth} and
  {Krautter}}}{1983}]{reipurth83}
{Reipurth} B. and {Krautter} J., 1983 \emph{\iaucirc}, \emph{3823}, 1.

\bibitem[\protect\astroncite{\emph{{Reipurth} et~al.}}{2002}]{reipurth02}
{Reipurth} B. et~al., 2002 \emph{\aj}, \emph{124}, 4, 2194.

\bibitem[\protect\astroncite{\emph{{Reipurth} et~al.}}{2012}]{reipurth12}
{Reipurth} B. et~al., 2012 \emph{\apjl}, \emph{748}, 1, L5.

\bibitem[\protect\astroncite{\emph{{Rice} et~al.}}{2015}]{rice15}
{Rice} T.~S. et~al., 2015 \emph{\aj}, \emph{150}, 4, 132.

\bibitem[\protect\astroncite{\emph{{Richards} et~al.}}{2012}]{richards12}
{Richards} J.~W. et~al., 2012 \emph{\apjs}, \emph{203}, 2, 32.

\bibitem[\protect\astroncite{\emph{{Rigliaco} et~al.}}{2020}]{rigliaco20}
{Rigliaco} E. et~al., 2020 \emph{\aap}, \emph{641}, A33.

\bibitem[\protect\astroncite{\emph{{Rigon} et~al.}}{2017}]{rigon17}
{Rigon} L. et~al., 2017 \emph{\mnras}, \emph{465}, 4, 3889.

\bibitem[\protect\astroncite{\emph{{Robinson} and
  {Espaillat}}}{2019}]{robinson19}
{Robinson} C.~E. and {Espaillat} C.~C., 2019 \emph{\apj}, \emph{874}, 2, 129.

\bibitem[\protect\astroncite{\emph{{Robinson} et~al.}}{2021}]{robinson21}
{Robinson} C.~E. et~al., 2021 \emph{\apj}, \emph{908}, 1, 16.

\bibitem[\protect\astroncite{\emph{{Rodriguez} and
  {Hillenbrand}}}{2022}]{rodriguez21}
{Rodriguez} A.~C. and {Hillenbrand} L.~A., 2022 \emph{\apj}, \emph{927}, 144.

\bibitem[\protect\astroncite{\emph{{Rodriguez} et~al.}}{2017}]{rodriguez17}
{Rodriguez} J.~E. et~al., 2017 \emph{\apj}, \emph{848}, 2, 97.

\bibitem[\protect\astroncite{\emph{{Roggero} et~al.}}{2021}]{roggero21}
{Roggero} N. et~al., 2021 \emph{\aap}, \emph{651}, A44.

\bibitem[\protect\astroncite{\emph{{Romanova} et~al.}}{2012}]{romanova12}
{Romanova} M.~M. et~al., 2012 \emph{\mnras}, \emph{421}, 1, 63.

\bibitem[\protect\astroncite{\emph{{Rosotti} et~al.}}{2017}]{rosotti17}
{Rosotti} G.~P. et~al., 2017 \emph{\mnras}, \emph{468}, 2, 1631.

\bibitem[\protect\astroncite{\emph{{Safron} et~al.}}{2015}]{safron15}
{Safron} E.~J. et~al., 2015 \emph{\apjl}, \emph{800}, L5.

\bibitem[\protect\astroncite{\emph{{S{\'a}nchez-S{\'a}ez}
  et~al.}}{2021}]{sanchez21}
{S{\'a}nchez-S{\'a}ez} P. et~al., 2021 \emph{\aj}, \emph{161}, 3, 141.

\bibitem[\protect\astroncite{\emph{{Scholz} et~al.}}{2013}]{scholz13}
{Scholz} A. et~al., 2013 \emph{\mnras}, \emph{430}, 4, 2910.

\bibitem[\protect\astroncite{\emph{{Semkov} et~al.}}{2010}]{semkov10}
{Semkov} E.~H. et~al., 2010 \emph{\aap}, \emph{523}, L3.

\bibitem[\protect\astroncite{\emph{{Sergison} et~al.}}{2020}]{sergison20}
{Sergison} D.~J. et~al., 2020 \emph{\mnras}, \emph{491}, 4, 5035.

\bibitem[\protect\astroncite{\emph{{Sicilia-Aguilar}
  et~al.}}{2012}]{sicilia-aguilar12}
{Sicilia-Aguilar} A. et~al., 2012 \emph{\aap}, \emph{544}, A93.

\bibitem[\protect\astroncite{\emph{{Sicilia-Aguilar}
  et~al.}}{2015}]{sicilia-aguilar15}
{Sicilia-Aguilar} A. et~al., 2015 \emph{\aap}, \emph{580}, A82.

\bibitem[\protect\astroncite{\emph{{Sicilia-Aguilar} et~al.}}{2017}]{sa17}
{Sicilia-Aguilar} A. et~al., 2017 \emph{\aap}, \emph{607}, A127.

\bibitem[\protect\astroncite{\emph{{Sicilia-Aguilar}
  et~al.}}{2020}]{sicilia-aguilar20}
{Sicilia-Aguilar} A. et~al., 2020 \emph{\aap}, \emph{643}, A29.

\bibitem[\protect\astroncite{\emph{{Siess} et~al.}}{2000}]{siess00}
{Siess} L. et~al., 2000 \emph{\aap}, \emph{358}, 593.

\bibitem[\protect\astroncite{\emph{{Sipos} et~al.}}{2009}]{sipos09}
{Sipos} N. et~al., 2009 \emph{\aap}, \emph{507}, 2, 881.

\bibitem[\protect\astroncite{\emph{{Siwak} et~al.}}{2018}]{siwak18}
{Siwak} M. et~al., 2018 \emph{\aap}, \emph{618}, A79.

\bibitem[\protect\astroncite{\emph{{Siwak} et~al.}}{2020}]{siwak20}
{Siwak} M. et~al., 2020 \emph{\aap}, \emph{644}, A135.

\bibitem[\protect\astroncite{\emph{{Sousa} et~al.}}{2016}]{sousa16}
{Sousa} A.~P. et~al., 2016 \emph{\aap}, \emph{586}, A47.

\bibitem[\protect\astroncite{\emph{{Sousa} et~al.}}{2021}]{sousa21}
{Sousa} A.~P. et~al., 2021 \emph{\aap}, \emph{649}, A68.

\bibitem[\protect\astroncite{\emph{{Stauffer} et~al.}}{2014}]{stauffer14}
{Stauffer} J. et~al., 2014 \emph{\aj}, \emph{147}, 4, 83.

\bibitem[\protect\astroncite{\emph{{Stauffer} et~al.}}{2015}]{stauffer15}
{Stauffer} J. et~al., 2015 \emph{\aj}, \emph{149}, 4, 130.

\bibitem[\protect\astroncite{\emph{{Stauffer} et~al.}}{2016}]{stauffer16}
{Stauffer} J. et~al., 2016 \emph{\aj}, \emph{151}, 3, 60.

\bibitem[\protect\astroncite{\emph{{Stecklum} et~al.}}{2021}]{stecklum21}
{Stecklum} B. et~al., 2021 \emph{\aap}, \emph{646}, A161.

\bibitem[\protect\astroncite{\emph{{Sugiyama} et~al.}}{2019}]{sugiyama19}
{Sugiyama} K. et~al., 2019 \emph{The Astronomer's Telegram}, \emph{12446}, 1.

\bibitem[\protect\astroncite{\emph{{Szab{\'o}} et~al.}}{2021}]{szabo21}
{Szab{\'o}} Z.~M. et~al., 2021 \emph{\apj}, \emph{917}, 2, 80.

\bibitem[\protect\astroncite{\emph{{Szab{\'o}} et~al.}}{2022}]{szabo22}
{Szab{\'o}} Z.~M. et~al., 2022 \emph{\apj}, \emph{936}, 1, 64.

\bibitem[\protect\astroncite{\emph{{Szegedi-Elek}
  et~al.}}{2020}]{szegedi-elek20}
{Szegedi-Elek} E. et~al., 2020 \emph{\apj}, \emph{899}, 2, 130.

\bibitem[\protect\astroncite{\emph{{Takasao} et~al.}}{2019}]{takasao19}
{Takasao} S. et~al., 2019 \emph{\apjl}, \emph{878}, 1, L10.

\bibitem[\protect\astroncite{\emph{{Taquet} et~al.}}{2016}]{taquet16}
{Taquet} V. et~al., 2016 \emph{\apj}, \emph{821}, 1, 46.

\bibitem[\protect\astroncite{\emph{{Tobin} et~al.}}{2020}]{tobin20}
{Tobin} J.~J. et~al., 2020 \emph{\apj}, \emph{905}, 2, 162.

\bibitem[\protect\astroncite{\emph{{Tofflemire} et~al.}}{2017}]{tofflemire17}
{Tofflemire} B.~M. et~al., 2017 \emph{\apjl}, \emph{842}, 2, L12.

\bibitem[\protect\astroncite{\emph{{Trimble}}}{1976}]{trimble76}
{Trimble} V., 1976 \emph{\qjras}, \emph{17}, 25.

\bibitem[\protect\astroncite{\emph{{Uchiyama} and
  {Ichikawa}}}{2019}]{uchiyama19}
{Uchiyama} M. and {Ichikawa} K., 2019 \emph{\apj}, \emph{883}, 1, 6.

\bibitem[\protect\astroncite{\emph{{van Roestel} et~al.}}{2021}]{vanroestel21}
{van Roestel} J. et~al., 2021 \emph{\aj}, \emph{161}, 6, 267.

\bibitem[\protect\astroncite{\emph{{van 't Hoff} et~al.}}{2018}]{hoff18}
{van 't Hoff} M. L.~R. et~al., 2018 \emph{\apjl}, \emph{864}, 1, L23.

\bibitem[\protect\astroncite{\emph{{Vazzano} et~al.}}{2021}]{vazzano21}
{Vazzano} M.~M. et~al., 2021 \emph{\aap}, \emph{648}, A41.

\bibitem[\protect\astroncite{\emph{{Venuti} et~al.}}{2015}]{venuti15}
{Venuti} L. et~al., 2015 \emph{\aap}, \emph{581}, A66.

\bibitem[\protect\astroncite{\emph{{Venuti} et~al.}}{2021}]{venuti21}
{Venuti} L. et~al., 2021 \emph{\aj}, \emph{162}, 3, 101.

\bibitem[\protect\astroncite{\emph{{Visser} et~al.}}{2015}]{visser15}
{Visser} R. et~al., 2015 \emph{\aap}, \emph{577}, A102.

\bibitem[\protect\astroncite{\emph{{Vorobyov} et~al.}}{2018}]{vorobyov18}
{Vorobyov} E.~I. et~al., 2018 \emph{\aap}, \emph{613}, A18.

\bibitem[\protect\astroncite{\emph{{Vorobyov} et~al.}}{2021}]{vorobyov21}
{Vorobyov} E.~I. et~al., 2021 \emph{\aap}, \emph{647}, A44.

\bibitem[\protect\astroncite{\emph{{Wenzel} and {Gessner}}}{1975}]{wenzel75}
{Wenzel} W. and {Gessner} H., 1975 \emph{Zentralinstitut fuer Astrophysik
  Sternwarte Sonneberg Mitteilungen ueber Veraenderliche Sterne}, \emph{7}, 23.

\bibitem[\protect\astroncite{\emph{{Wolk} et~al.}}{2015}]{wolk15}
{Wolk} S.~J. et~al., 2015 \emph{\aj}, \emph{150}, 5, 145.

\bibitem[\protect\astroncite{\emph{{Wolk} et~al.}}{2018}]{wolk18}
{Wolk} S.~J. et~al., 2018 \emph{\aj}, \emph{155}, 2, 99.

\bibitem[\protect\astroncite{\emph{{Yoo} et~al.}}{2017}]{yoo17}
{Yoo} H. et~al., 2017 \emph{\apj}, \emph{849}, 69.

\bibitem[\protect\astroncite{\emph{{Yoon} et~al.}}{2022}]{yoon22}
{Yoon} S.-Y. et~al., 2022 \emph{\apj}, \emph{929}, 1, 60.

\bibitem[\protect\astroncite{\emph{{Zakri} et~al.}}{2022}]{zakri22}
{Zakri} W. et~al., 2022 \emph{\apjl}, \emph{924}, 2, L23.

\bibitem[\protect\astroncite{\emph{{Zhang} et~al.}}{2019}]{zhang19}
{Zhang} Y. et~al., 2019 \emph{\apj}, \emph{883}, 1, 1.

\bibitem[\protect\astroncite{\emph{{Zhu} et~al.}}{2009}]{zhu09}
{Zhu} Z. et~al., 2009 \emph{\apj}, \emph{694}, 2, 1045.

\bibitem[\protect\astroncite{\emph{{Zhu} et~al.}}{2010}]{zhu10}
{Zhu} Z. et~al., 2010 \emph{\apj}, \emph{713}, 2, 1134.

\bibitem[\protect\astroncite{\emph{{Zsidi} et~al.}}{2019}]{zsidi19}
{Zsidi} G. et~al., 2019 \emph{\apj}, \emph{873}, 2, 130.

\bibitem[\protect\astroncite{\emph{{Zsidi} et~al.}}{2022}]{zsidi22}
{Zsidi} G. et~al., 2022 \emph{\aap}, \emph{660}, A108.

\end{thebibliography}
}

\end{document}